\documentclass[preprint,12pt]{elsarticle}
\usepackage[margin=1in]{geometry}
\usepackage{graphicx}
\usepackage{amsmath}
\usepackage{booktabs}
\usepackage{hyperref}
\usepackage{comment}
\journal{}

\begin{document}

\begin{frontmatter}

\title{Inherent Failure Modes of Physics-Informed Neural Networks for Buoyancy-Driven Flow: Role of Non-Dimensionalization in Network Training}

\author[K. Chand]{K. Chand\corref{cor1}}
\author[A. Peyvan]{A. Peyvan}
\author[H. Luo]{H. Luo}

\cortext[cor1]{Corresponding author. Email: krishan.chand@vanderbilt.edu}

\address{Department of Mechanical Engineering, Vanderbilt University, 2201 West End Avenue, Nashville, TN 37235, USA}


\begin{abstract}
In this work, we demonstrate that non-dimensionalization plays a critical role in the training and accuracy of physics-informed neural networks (PINNs) for buoyancy-driven flows. We investigate the influence of two velocity scalings, thermal-diffusive and free-fall, on PINN predictions for a differentially heated square cavity at a fixed Prandtl number of $\mathrm{Pr}=0.71$ and Rayleigh numbers ranging from $\mathrm{Ra}=10^3$ to $10^5$. In the thermal-diffusive scaling, the buoyancy term carries a coefficient $B=\mathrm{Ra}\,\mathrm{Pr}$, whereas the free-fall scaling yields $B=1$. Under identical network architectures and training conditions, the free-fall PINNs exhibit smooth convergence and agree with the reference direct numerical simulation (DNS) within 1.1\% in the mean Nusselt number ($\overline{\mathrm{Nu}}$) for $\mathrm{Ra}\le10^4$, with an error of 11.5\% at $\mathrm{Ra}=10^5$. In contrast, the diffusive PINNs exhibit errors of 157--650\% at $\mathrm{Ra}\le10^4$ and diverge at $\mathrm{Ra}=10^5$. We attribute this difference to a coefficient-induced gradient imbalance, for which the dominant buoyancy-related contribution to the loss gradient scales as $(\mathrm{Ra}\,\mathrm{Pr})^2$. Direct measurements of the individual loss gradients confirm a severe imbalance in the diffusive formulation. Furthermore, two commonly used mitigation strategies, sparse-data supervision and gradient-norm-based adaptive loss weighting, fail to recover accurate solutions under the tested conditions. We also extend the free-fall formulation to unsteady convection developing from rest at $\mathrm{Ra}=10^4$ using two random initializations. Although the predicted long-time mean Nusselt number agrees with the reference within 0.7\%, the PINNs fail to capture the transient overshoot (reference peak 2.41 versus approximately 2.30 for the PINNs) and exhibit velocity errors of approximately 20\% during the early transient. These findings highlight the importance of selecting an appropriate non-dimensionalization to improve PINN trainability while revealing additional challenges in accurately resolving transient convection.
\end{abstract}


\end{frontmatter}

\section{Introduction}
\label{sec:intro}
Physics-informed neural networks (PINNs) provide a framework for approximating solutions to partial differential equations (PDEs), particularly when observational data are limited. The network is trained by minimizing a loss function comprising the residuals of the governing equations evaluated at collocation points, together with the boundary- and initial-condition constraints, without requiring a computational mesh. PINNs were introduced in their modern form by Raissi et al.\ \citep{raissi2019} as a general framework for solving forward and inverse problems involving nonlinear PDEs. Since observational data can also be incorporated into the loss function, PINNs have attracted considerable interest for inverse problems and data assimilation. The framework has been implemented in general-purpose software such as DeepXDE \citep{lu2021deepxde} and applied to a wide range of problems in fluid mechanics, heat transfer, and solid mechanics \citep{karniadakis2021}.

Despite their flexibility, PINNs often encounter significant difficulties during training, commonly referred to as failure modes. One major challenge arises from the imbalance among the different contributions to the loss function, including the PDE residuals and boundary- and initial-condition terms, whose magnitudes and gradients can differ substantially. Wang et al.\ \citep{wang2021} investigated this issue through neural tangent kernel analysis and demonstrated that imbalanced gradients can lead to poor convergence of individual loss components. They proposed a learning-rate annealing strategy that adaptively adjusts the loss weights based on gradient statistics. McClenny and Braga-Neto \citep{mcclenny2023} introduced self-adaptive weights at individual collocation points, allowing the network to emphasize regions that are more difficult to satisfy. Another approach is to enforce boundary conditions directly through the network architecture, using trial-function formulations \citep{lagaris1998} or approximate distance functions \citep{sukumar2022}. This eliminates the corresponding boundary-condition penalties from the loss function. Although these approaches address different aspects of PINN optimization, they primarily operate on the loss formulation after the governing equations have been specified. However, the coefficients introduced by the non-dimensionalization of the governing equations can themselves influence the relative magnitudes of the residuals and their gradients, potentially introducing an additional source of training imbalance.

PINNs have also been applied to a variety of natural- and turbulent-convection problems. Cai et al.\ \citep{cai2021heat} reviewed PINN applications to heat transfer, including convection--diffusion problems, and highlighted their potential for inverse problems involving sparse temperature and velocity measurements. Lucor et al.\ \citep{lucor2022} developed PINN surrogates for turbulent Rayleigh--B\'{e}nard convection using sparse DNS data and demonstrated that padding and normalization strategies can improve accuracy near boundaries. Fowler et al.\ \citep{fowler2024} investigated the two-dimensional differentially heated cavity using a vorticity--streamfunction formulation and validated their predictions against established benchmarks over a range of Rayleigh numbers. These studies demonstrate the potential of PINNs to reproduce buoyancy-driven flows through suitable network architectures, loss formulations, and, in some cases, auxiliary data. However, the influence of non-dimensionalization on PINN training, particularly under otherwise identical network architectures and training conditions, remains insufficiently explored.

The differentially heated square cavity provides a well-established benchmark for investigating buoyancy-driven flows. De Vahl Davis \citep{devahldavis1983} reported benchmark solutions at $\mathrm{Pr}=0.71$ over $\mathrm{Ra}=10^3$--$10^6$, while Wan et al.\ \citep{wan2001} extended the benchmark to $\mathrm{Ra}=10^8$ using two independent high-accuracy numerical schemes. Wan et al.\ also emphasized that comparisons based solely on global scalar quantities, such as the maximum velocity or mean Nusselt number, may be misleading because different flow structures can produce similar values of these quantities. Therefore, the present study evaluates both global quantities and full-field predictions against an independently computed DNS reference and the available benchmark results of Wan et al.\ \citep{wan2001}.

The Boussinesq equations governing buoyancy-driven flow can be expressed using different characteristic velocity scales without changing the underlying physics. In the conventional diffusive scaling, the characteristic velocity is defined as $U=\alpha/L$, resulting in a buoyancy coefficient $B=\mathrm{Ra}\,\mathrm{Pr}$ in the momentum equation. In contrast, the free-fall scaling, based on $U=\sqrt{g\beta\Delta T L}$, yields a unit buoyancy coefficient, $B=1$, while the viscous and thermal diffusion coefficients become $\sqrt{\mathrm{Pr}/\mathrm{Ra}}$ and $1/\sqrt{\mathrm{Ra}\,\mathrm{Pr}}$, respectively. Although these two non-dimensionalizations represent the same physical system, their governing equations contain substantially different coefficient magnitudes. Consequently, their residuals and associated gradients can differ significantly during PINN optimization, even when the network architecture and training procedure remain unchanged. This raises an important question: to what extent does the choice of non-dimensionalization influence the convergence and accuracy of PINNs for buoyancy-driven flows?

To address this question, we systematically compare diffusive and free-fall scalings for the differentially heated cavity over $\mathrm{Ra}=10^3$--$10^5$, while keeping the network architecture, collocation-point budget, optimizer, and training iterations fixed. We demonstrate that the diffusive PINNs exhibit large errors in the mean Nusselt number, ranging from approximately 158\% to 649\%, and eventually diverge at the highest Rayleigh number. In contrast, the free-fall PINNs exhibit substantially better convergence and agreement with DNS, particularly at the two lower Rayleigh numbers. We further investigate the underlying mechanism by analyzing the gradients of individual loss terms and show that the buoyancy coefficient introduces a gradient contribution proportional to $B^2$ under the assumptions considered. We also evaluate whether sparse data assimilation and adaptive loss weighting can mitigate the resulting training difficulties. Finally, we extend the free-fall formulation to unsteady convection developing from rest to assess its ability to predict the temporal evolution of heat transfer and flow structures.

The present study highlights non-dimensionalization as an important consideration in PINN optimization, in addition to network architecture, collocation sampling, and loss weighting. Unlike strategies that modify the relative contributions of individual loss terms, an appropriate non-dimensionalization can reduce coefficient-induced gradient amplification directly within the governing residuals. Although the present investigation focuses on buoyancy-driven convection, similar considerations may be relevant to other multiphysics systems involving strongly varying source or forcing coefficients, such as rotating flows where Coriolis terms enter the momentum equations \citep{chand_turbulent_2025}. The findings therefore provide a physical and numerical basis for selecting characteristic scales when developing PINNs for coupled transport problems.

The remainder of the paper is organized as follows. Section~\ref{sec:formulation} presents the governing equations under both non-dimensionalizations and the DNS reference solution. Section~\ref{sec:methods} describes the PINN architecture, boundary-condition enforcement, training procedure, sparse-data assimilation, adaptive loss weighting, and unsteady formulation. Section~\ref{sec:results} presents the training convergence, accuracy comparisons with DNS, and predicted flow structures. The mechanism underlying gradient imbalance is examined in Sec.~\ref{sec:discussion}, followed by an evaluation of the mitigation strategies and unsteady predictions. Finally, the main conclusions are summarized in Sec.~\ref{sec:conclusion}.

\section{Governing equations and non-dimensionalization}
\label{sec:formulation}

Natural convection in a differentially heated square cavity of side length $L$ is governed by the incompressible Navier--Stokes equations under the Boussinesq approximation. The dimensional governing equations are given by

\begin{align}
\bar u_{\bar x} + \bar v_{\bar y} &= 0, \label{eq:dim-cont} \\
\rho_0\bar{u}_t+\rho_0\!\left(\bar u \bar u_{\bar x} + \bar v \bar u_{\bar y}\right) &= -\bar p_{\bar x} + \mu\!\left(\bar u_{\bar x\bar x}+\bar u_{\bar y\bar y}\right), \label{eq:dim-momx} \\
\rho_0\bar{v}_t + \rho_0\!\left(\bar u \bar v_{\bar x} + \bar v \bar v_{\bar y}\right) &= -\bar p_{\bar y} + \mu\!\left(\bar v_{\bar x\bar x}+\bar v_{\bar y\bar y}\right)  + \rho_0 g \beta (\bar\theta - \bar\theta_{\mathrm{ref}}), \label{eq:dim-momy} \\
\bar{\theta}_t + \bar u \bar\theta_{\bar x} + \bar v \bar\theta_{\bar y} &= \alpha\!\left(\bar\theta_{\bar x\bar x}+\bar\theta_{\bar y\bar y}\right), \label{eq:dim-energy}
\end{align}

where $\bar u$, $\bar v$, and $\bar\theta$ denote the dimensional horizontal velocity, vertical velocity, and temperature, respectively. The overbar indicates dimensional quantities, while subscripts denote partial derivatives. Here, $\rho_0$ is the reference density, $\mu$ is the dynamic viscosity, $\alpha$ is the thermal diffusivity, $\beta$ is the thermal expansion coefficient, and $g$ is the magnitude of gravitational acceleration acting in the negative vertical direction.

\begin{figure}[t!]
\centering
\includegraphics[width=0.4\textwidth]{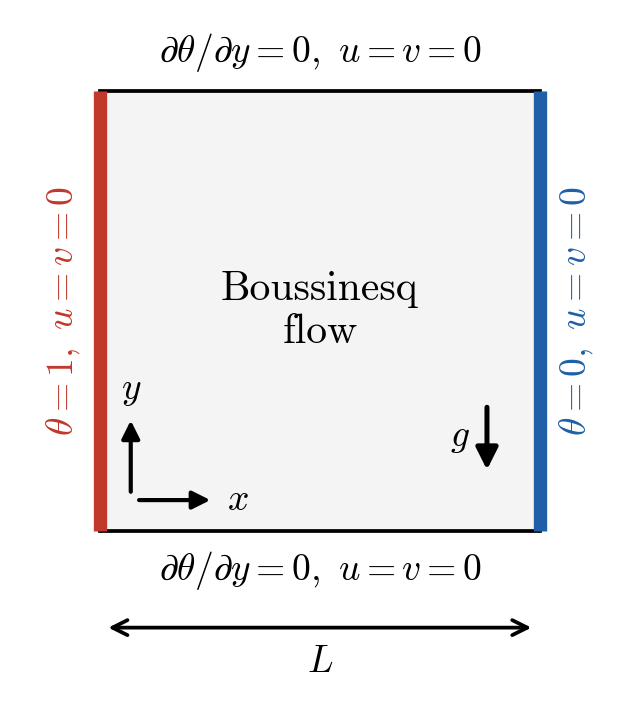}
\caption{Schematic of the differentially heated square cavity and its boundary conditions. The left and right vertical walls are maintained at constant temperatures $\theta=1$ and $\theta=0$, respectively, while the horizontal walls are adiabatic. No-slip conditions are imposed on all walls, and gravity acts in the negative $y$-direction.}
\label{fig:schematic}
\end{figure}

For both non-dimensionalizations considered in this study, the spatial coordinates and temperature are defined as

\[
x=\frac{\bar x}{L},\qquad
y=\frac{\bar y}{L},\qquad
\theta=\frac{\bar\theta-\bar\theta_C}{\bar\theta_H-\bar\theta_C},
\]

where $\bar\theta_H$ and $\bar\theta_C$ are the temperatures of the hot and cold walls, respectively. The two formulations differ in their characteristic velocity scales and, consequently, their time and pressure scales. Figure~\ref{fig:schematic} illustrates the cavity geometry and boundary conditions.

\subsection{Thermal-diffusive scaling}
\label{sec:diffusive-scaling}

In the thermal-diffusive scaling, the characteristic velocity and time scales are defined as $U=\alpha/L$ and $L^2/\alpha$, respectively. The resulting non-dimensional governing equations take the conventional form used by de Vahl Davis \citep{devahldavis1983} and Wan et al.\ \citep{wan2001}:

\begin{align}
u_x + v_y &= 0, \label{eq:cont} \\
u_t + u u_x + v u_y &= -p_x + \mathrm{Pr}\,(u_{xx}+u_{yy}), \label{eq:diff-momx} \\
v_t + u v_x + v v_y &= -p_y + \mathrm{Pr}\,(v_{xx}+v_{yy}) + \mathrm{Ra}\,\mathrm{Pr}\,\theta, \label{eq:diff-momy} \\
\theta_t + u \theta_x + v \theta_y &= \theta_{xx}+\theta_{yy}, \label{eq:diff-energy}
\end{align}

where the Rayleigh and Prandtl numbers are defined as

\[
\mathrm{Ra}=\frac{g\beta\Delta\theta L^3}{\nu\alpha},
\qquad
\mathrm{Pr}=\frac{\nu}{\alpha},
\]

with $\Delta\theta=\bar\theta_H-\bar\theta_C$ and $\nu=\mu/\rho_0$ denoting the temperature difference and kinematic viscosity, respectively.

In this formulation, the buoyancy coefficient in the vertical momentum equation is $B=\mathrm{Ra}\,\mathrm{Pr}$. For the Rayleigh numbers considered here, this coefficient varies from 710 to $7.1\times10^4$, as summarized in Table~\ref{tab:coefficients}. Thus, the buoyancy contribution becomes increasingly large with increasing $Ra$, potentially affecting the relative magnitudes of the PDE residuals during PINN training.

\subsection{Free-fall scaling}

An alternative non-dimensionalization is based on the free-fall velocity scale, $U=\sqrt{g\beta\Delta\theta L}$, with the corresponding time scale $L/U$. The governing equations then become

\begin{align}
u_x + v_y &= 0, \label{eq:ff-cont} \\
u_t + u u_x + v u_y &= -p_x + \sqrt{\mathrm{Pr}/\mathrm{Ra}}\,(u_{xx}+u_{yy}), \label{eq:ff-momx} \\
v_t + u v_x + v v_y &= -p_y + \sqrt{\mathrm{Pr}/\mathrm{Ra}}\,(v_{xx}+v_{yy}) + \theta, \label{eq:ff-momy} \\
\theta_t + u \theta_x + v \theta_y &= \tfrac{1}{\sqrt{\mathrm{Ra}\,\mathrm{Pr}}}\,(\theta_{xx}+\theta_{yy}). \label{eq:ff-energy}
\end{align}

Unlike the diffusive formulation, the free-fall scaling yields a unit buoyancy coefficient, $B=1$, independent of $Ra$ and $Pr$. The viscous and thermal diffusion coefficients are $\sqrt{\mathrm{Pr}/\mathrm{Ra}}$ and $1/\sqrt{\mathrm{Ra}\,\mathrm{Pr}}$, respectively, both of which decrease with increasing $Ra$ at fixed $Pr$ (Table~\ref{tab:coefficients}).

Although the two formulations describe the same physical system, their non-dimensional velocity, streamfunction, and pressure fields differ because of the characteristic scales. In particular, the diffusive velocity and streamfunction are $\sqrt{\mathrm{Ra}\,\mathrm{Pr}}$ times their free-fall counterparts, while the corresponding pressure fields differ by a factor of $\mathrm{Ra}\,\mathrm{Pr}$. In contrast, the non-dimensional temperature and Nusselt number are independent of the velocity scaling. These relationships allow direct comparisons between the two formulations after the appropriate conversions.

\subsection{DNS reference}
\label{sec:dns}

To assess the accuracy of the PINN predictions, we perform independent direct numerical simulations (DNS) at all three Rayleigh numbers using an in-house finite-volume solver based on a collocated grid \citep{chand_geometry-controlled_2026,chand_statistics_2019}. The advection terms are advanced explicitly using the Adams--Bashforth scheme, while the diffusive terms are treated using the Crank--Nicolson scheme. Further details of the numerical formulation and solver implementation are available in Refs.~\citep{sharma_dominant_2024,sharma_influence_2022,chand_significance_2024}. The time step is selected to maintain a Courant--Friedrichs--Lewy (CFL) number below 0.3.

The DNS is performed using the free-fall formulation given by Eqs.~\eqref{eq:ff-cont}--\eqref{eq:ff-energy}. For direct comparison with the diffusive PINNs, the DNS velocity fields are converted to the diffusive scale by multiplying the free-fall velocities by $\sqrt{\mathrm{Ra}\,\mathrm{Pr}}$. The temperature and Nusselt number require no conversion because they are independent of the velocity scaling.

A uniform $64\times64$ collocated grid is employed for all three Rayleigh numbers, providing the spatial resolution used for the present comparisons \citep{chand_pressure-free_2024}. Since the trained PINN represents a continuous and differentiable approximation of the solution, its predictions can be evaluated directly at the DNS grid locations without requiring interpolation of the PINN fields. This enables a consistent pointwise comparison between the PINN and DNS results.

\begin{table}[t!]
\centering
\caption{Coefficients of the viscous, thermal diffusion, and buoyancy terms in the non-dimensional governing equations for diffusive and free-fall scalings at $\mathrm{Pr}=0.71$. The buoyancy coefficient increases by two orders of magnitude over the considered Rayleigh-number range in the diffusive formulation, whereas it remains unity in the free-fall formulation.}
\label{tab:coefficients}
\begin{tabular}{@{}llccc@{}}
\toprule
Scaling & $\mathrm{Ra}$ & Viscosity & Diffusivity & Buoyancy \\
\midrule
Diffusive & $10^3$ & 0.71 & 1 & 710 \\
Diffusive & $10^4$ & 0.71 & 1 & 7{,}100 \\
Diffusive & $10^5$ & 0.71 & 1 & 71{,}000 \\
\addlinespace
Free-fall & $10^3$ & 0.0267 & 0.0375 & 1 \\
Free-fall & $10^4$ & 0.00843 & 0.0119 & 1 \\
Free-fall & $10^5$ & 0.00267 & 0.00375 & 1 \\
\bottomrule
\end{tabular}
\end{table}

\section{PINN methodology}
\label{sec:methods}

\begin{figure}[ht]
\centering
\includegraphics[width=0.9\textwidth]{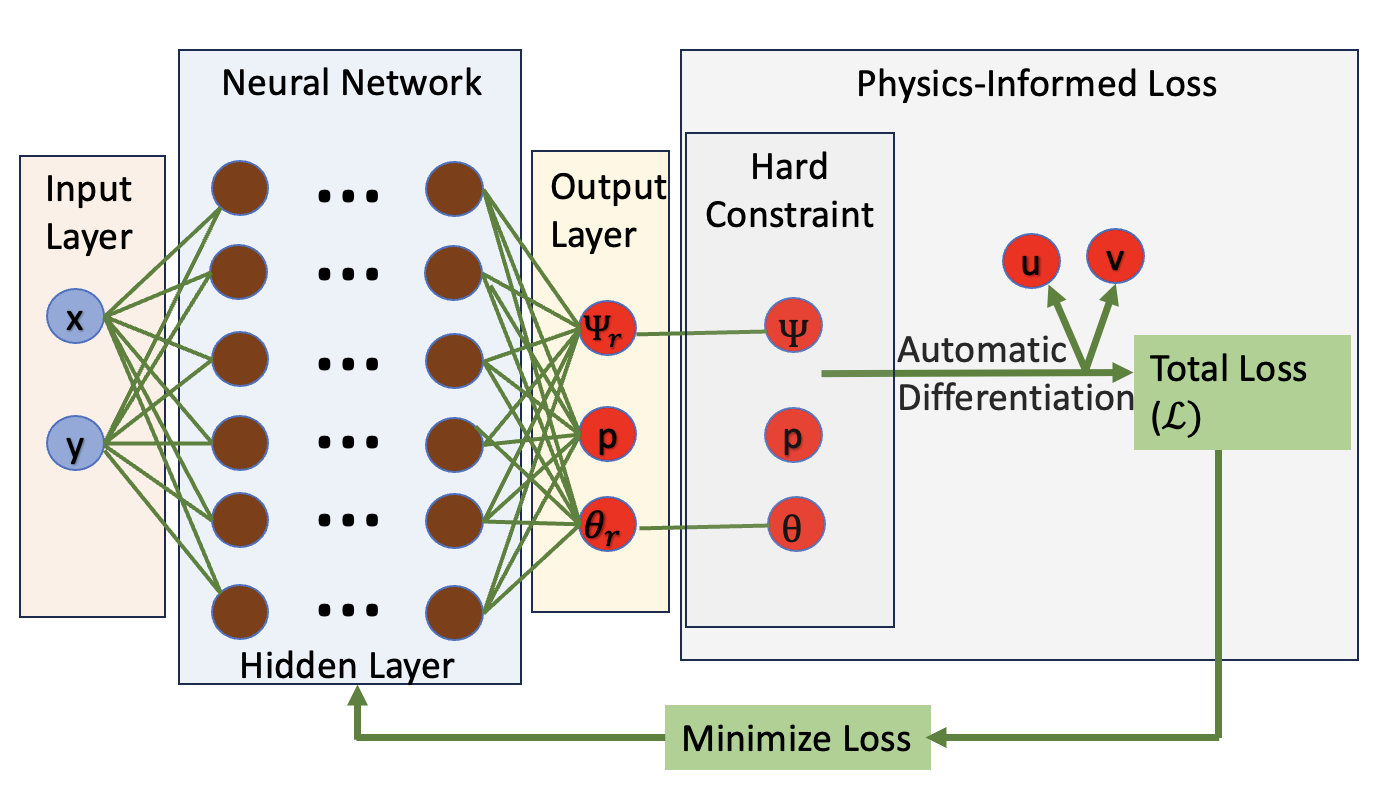}
\caption{Schematic of the PINN architecture comprising four hidden layers with 64 neurons each. The network predicts the streamfunction, pressure, and temperature fields. Boundary conditions are enforced through a combination of hard constraints, defined in Eq.~\ref{eq:constraint}, and soft constraints incorporated into the loss function.}
\label{fig:schematic_nn}
\end{figure}

\subsection{Network architecture}
\label{sec:architecture}

As illustrated in Fig.~\ref{fig:schematic_nn}, we employ a fully connected multilayer perceptron consisting of two spatial inputs ($x,y$), four hidden layers with 64 neurons each, and three outputs $(\psi_{\mathrm{raw}},p,\theta_{\mathrm{raw}})$, where $\psi_{\mathrm{raw}}$ and $\theta_{\mathrm{raw}}$ are the raw streamfunction and temperature outputs, respectively. Each hidden layer uses the hyperbolic tangent ($\tanh$) activation function, with Xavier-normal weight initialization and zero biases.

The choice of a smooth activation function is important because the PDE residuals require higher-order derivatives of the network outputs, which are computed using automatic differentiation. In contrast, piecewise-linear activation functions such as ReLU have vanishing second derivatives almost everywhere and are therefore generally unsuitable for directly representing the required viscous and diffusive terms.

The velocity components are recovered from the streamfunction as

\[
u=\frac{\partial\psi}{\partial y},
\qquad
v=-\frac{\partial\psi}{\partial x}.
\]

For a sufficiently smooth streamfunction, the continuity equation is satisfied identically:

\[
u_x+v_y=\psi_{yx}-\psi_{xy}=0.
\]

Consequently, the continuity residual does not need to be included explicitly in the training loss.

\subsection{Hard-constrained boundary conditions}

As shown in Fig.~\ref{fig:schematic_nn}, the impermeability condition and the prescribed temperatures at the vertical walls are enforced directly through the network outputs using the following transformations:

\begin{equation}\label{eq:constraint}
\psi=x(1-x)\,y(1-y)\,\psi_{\mathrm{raw}},
\qquad
\theta=(1-x)+x(1-x)\,\theta_{\mathrm{raw}}.
\end{equation}

These transformations ensure that $\psi=0$ along all four walls and that $\theta=1$ and $\theta=0$ at the hot ($x=0$) and cold ($x=1$) walls, respectively, regardless of the raw network predictions. The constant streamfunction along the cavity boundary enforces the no-penetration condition. However, the tangential no-slip condition is not automatically satisfied by this construction. Similarly, the adiabatic conditions at the horizontal walls, $\theta_y=0$ at $y=0$ and $y=1$, are not enforced by the temperature transformation. These remaining conditions are therefore imposed as soft constraints through the loss function.

The use of hard constraints was motivated by preliminary training experiments. When the temperature Dirichlet conditions were imposed only through a soft penalty, the diffusive PINN approached a trivial solution characterized by $\theta\approx0$ and $u,v\approx0$. Such a state can produce small interior PDE residuals while failing to satisfy the prescribed thermal boundary conditions. Incorporating the temperature Dirichlet conditions directly into the network output prevents this particular failure mode. Nevertheless, as demonstrated in Sec.~\ref{sec:results}, hard boundary constraints alone are insufficient to ensure accurate convergence under diffusive scaling.

\subsection{Loss function and training}
\label{sec:loss}

The total training loss consists of the PDE residual loss ($\mathcal{L}_{\mathrm{pde}}$), the velocity boundary-condition loss ($\mathcal{L}_{\mathrm{vel}}$), the adiabatic boundary-condition loss ($\mathcal{L}_{\theta}$), and a pressure normalization term ($\mathcal{L}_{p}$). The combined loss function is defined as

\begin{equation}
\mathcal{L}
=
\mathcal{L}_{\mathrm{pde}}
+20\,\mathcal{L}_{\mathrm{vel}}
+20\,\mathcal{L}_{\theta}
+\mathcal{L}_{p}.
\label{eq:loss-total}
\end{equation}

Here, $\mathcal{L}_{\mathrm{pde}}$ represents the mean squared residuals of the two momentum equations and the energy equation, evaluated using either Eqs.~\eqref{eq:diff-momx}--\eqref{eq:diff-energy} or Eqs.~\eqref{eq:ff-momx}--\eqref{eq:ff-energy}. The PDE residuals are evaluated at 8,000 randomly distributed interior collocation points. The velocity boundary loss, $\mathcal{L}_{\mathrm{vel}}$, is computed from the mean squared velocity components at 400 points per wall, while $\mathcal{L}_{\theta}$ penalizes the mean squared temperature gradient $\theta_y$ at the same collocation points along the horizontal walls. The pressure normalization term, $\mathcal{L}_p=\langle p\rangle^2$, removes the arbitrary additive constant in the pressure field.

To isolate the influence of non-dimensionalization on PINN training, the network architecture, loss formulation, loss weights, collocation-point budget, and random seed are kept identical between the diffusive and free-fall cases at each Rayleigh number. All six configurations are trained using 5,000 iterations of the Adam optimizer, with an initial learning rate of $10^{-3}$ reduced to zero through cosine annealing, followed by 3,000 L-BFGS iterations using a strong-Wolfe line search. Thus, the principal variables in the comparison are the choice of non-dimensionalization and the Rayleigh number.

Additional details of the sparse-data assimilation and adaptive loss weighting strategies used to investigate gradient imbalance are provided in the Appendix.

\section{Results and discussion}
\label{sec:results}
In this section, we first examine the training convergence, followed by an assessment of the accuracy of the PINN predictions and the resulting flow structures. We also demonstrate that two well-known approaches for mitigating gradient imbalance in the loss function fail to resolve the training difficulties.

\begin{figure}[t!]
\centering
\includegraphics[width=\textwidth]{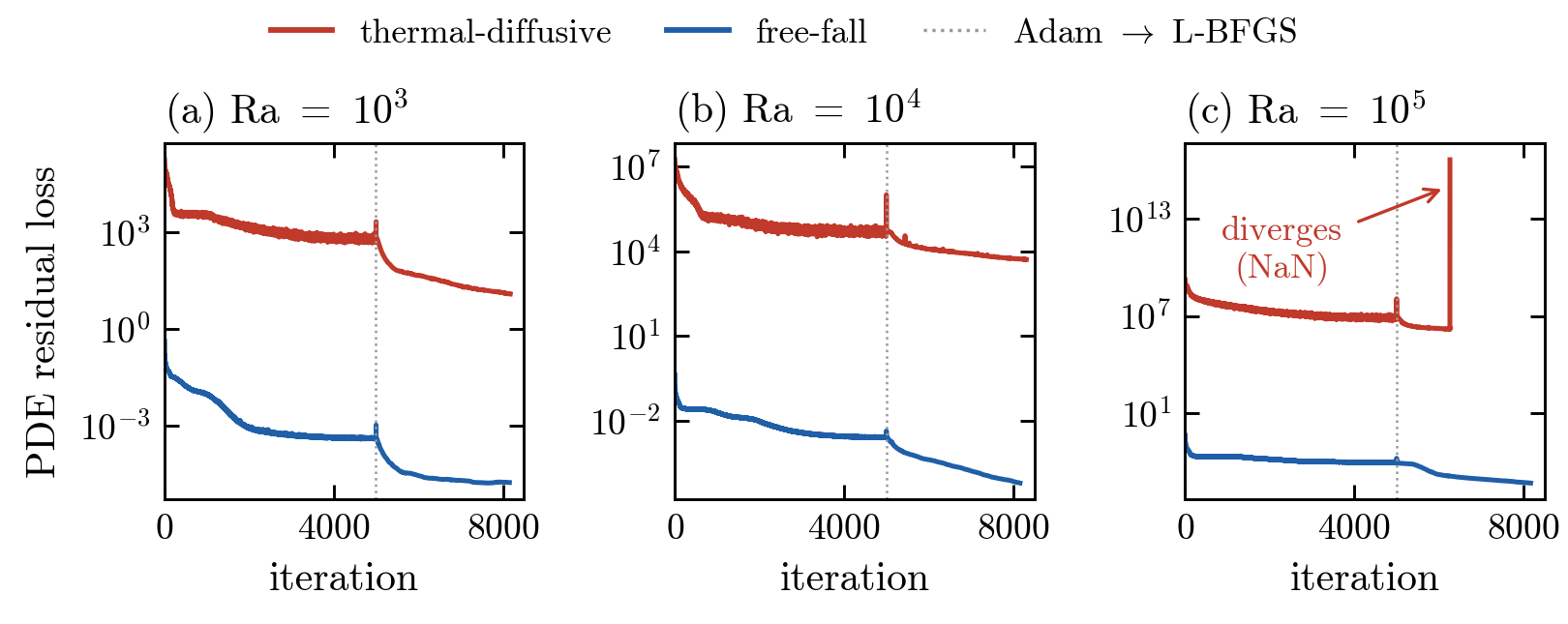}
\caption{Evolution of the PDE residual loss with training iteration for the diffusive and free-fall non-dimensionalizations at (a) $\mathrm{Ra}=10^3$, (b) $\mathrm{Ra}=10^4$, and (c) $\mathrm{Ra}=10^5$. Training is performed using Adam for the first 5,000 iterations, followed by L-BFGS. The diffusive PINN at $\mathrm{Ra}=10^5$ diverges during L-BFGS optimisation, resulting in a final loss of \texttt{NaN}.}
\label{fig:convergence}
\end{figure}
\subsection{Training convergence}

We begin by comparing the training convergence of the diffusive and free-fall PINNs. Figure~\ref{fig:convergence} shows the evolution of the PDE residual loss during training for both scalings at increasing Rayleigh numbers ($Ra$). For the free-fall scaling, the loss decreases smoothly and nearly monotonically, reaching final values of $1.8\times10^{-5}$, $6.4\times10^{-5}$, and $4.8\times10^{-4}$ at $\mathrm{Ra}=10^3$, $10^4$, and $10^5$, respectively. The spike observed at iteration 5,000 is associated with the switch from the Adam optimiser to L-BFGS. Furthermore, the initial loss increases with $Ra$, suggesting that higher Rayleigh numbers may require more training iterations to achieve comparable convergence.

In contrast, the loss for the diffusive scaling plateaus at values 3--8 orders of magnitude higher than those obtained with the free-fall scaling during Adam training across all three Rayleigh numbers. At $\mathrm{Ra}=10^5$, the loss increases abruptly during L-BFGS optimisation and eventually diverges, as indicated by the spike approaching $10^{16}$ in Fig.~\ref{fig:convergence}(c). For $\mathrm{Ra}=10^3$ and $10^4$, the diffusive PINNs also exhibit substantially higher final losses than their free-fall counterparts, with values of 12.1 and 5079, respectively. At $\mathrm{Ra}=10^5$, the final loss becomes \texttt{NaN}, indicating a complete failure of convergence.

\subsection{Accuracy comparison against DNS}

We further assess the accuracy of the PINN predictions by comparing them with direct numerical simulation (DNS) results. Table~\ref{tab:dns} presents the percentage errors in the mean Nusselt number ($\overline{\mathrm{Nu}}$) and the maximum horizontal and vertical velocities ($u_{\max}$ and $v_{\max}$). As $Ra$ increases from $10^3$ to $10^5$, the error in $\overline{\mathrm{Nu}}$ increases for both scalings. The free-fall PINNs show excellent agreement with DNS up to $\mathrm{Ra}=10^4$, with errors of 0.0\% and 1.1\% at $\mathrm{Ra}=10^3$ and $10^4$, respectively. In contrast, the diffusive PINNs exhibit substantially larger errors, reaching 157.7\% and 648.5\% at the corresponding Rayleigh numbers. At $\mathrm{Ra}=10^5$, the diffusive PINN fails to converge, whereas the free-fall PINN yields an error of 11.5\% in $\overline{\mathrm{Nu}}$. The relatively larger error in the latter case may be attributed to the fixed training budget used across all Rayleigh numbers, suggesting that additional training iterations may be required at higher $Ra$. As an independent validation of the DNS results, the computed value of $\overline{\mathrm{Nu}}=1.101$ at $\mathrm{Ra}=10^3$ agrees within 2.6\% with the value of 1.073 reported by Wan et al.\ \citep{wan2001}.

The errors in the maximum velocity components further highlight the differences between the two scalings. For the free-fall PINNs, the errors in $u_{\max}$ and $v_{\max}$ remain below 5\% across all three Rayleigh numbers. In contrast, the diffusive PINNs exhibit considerably larger errors, particularly at $\mathrm{Ra}=10^4$, where the errors reach 84.1\% and 54.2\% for $u_{\max}$ and $v_{\max}$, respectively. Although the diffusive PINNs show a decreasing loss trend at the two lower Rayleigh numbers, their substantially higher residual losses indicate that the solutions remain far from the DNS reference results within the prescribed training budget.

\begin{table}[t!]
\centering
\caption{Comparison of the mean Nusselt number and maximum velocity components between PINN predictions and DNS at different Rayleigh numbers. All quantities are evaluated on the $64\times64$ DNS grid, and percentage errors are calculated relative to DNS.}
\label{tab:dns}
\begin{tabular}{@{}llrrrr@{}}
\toprule
Scaling & $\mathrm{Ra}$ & $\overline{\mathrm{Nu}}$ (DNS) & $\overline{\mathrm{Nu}}$ err.\ (\%) & $u_{\max}$ err.\ (\%) & $v_{\max}$ err.\ (\%) \\
\midrule
Diffusive & $10^3$ & 1.101 & 157.7 & 1.9 & 20.3 \\
Free-fall & $10^3$ & 1.101 & 0.0 & 0.8 & 1.1 \\
\addlinespace
Diffusive & $10^4$ & 2.218 & 648.5 & 84.1 & 54.2 \\
Free-fall & $10^4$ & 2.218 & 1.1 & 0.3 & 0.3 \\
\addlinespace
Diffusive & $10^5$ & 4.492 & \texttt{NaN} & \texttt{NaN} & \texttt{NaN} \\
Free-fall & $10^5$ & 4.492 & 11.5 & 2.4 & 4.3 \\
\bottomrule
\end{tabular}
\end{table}

As noted by Wan et al.\ \citep{wan2001}, comparisons based solely on global scalar quantities may be misleading, as they do not necessarily capture discrepancies in the spatial structure of the solution. Therefore, we also evaluate the full-field relative $L_2$ error over the entire computational grid, defined as

\begin{equation}\label{eq:fullfield_relativeerror}
    \mathrm{L}_2 = \frac{\|\phi_{\mathrm{PINN}}-\phi_{\mathrm{DNS}}\|_2}{\|\phi_{\mathrm{DNS}}\|_2},
\end{equation}

where $\phi$ represents the velocity components ($u$, $v$) or temperature ($\theta$). Table~\ref{tab:fielderror} summarizes the corresponding errors for both scalings. The free-fall PINNs maintain relative errors below 1.5\% in all three fields up to $\mathrm{Ra}=10^4$, demonstrating close agreement with DNS. At $\mathrm{Ra}=10^5$, the errors increase to 20.9\%, 14.2\%, and 9.7\% for $u$, $v$, and $\theta$, respectively, indicating a deterioration in accuracy at higher Rayleigh numbers under the fixed training budget. In contrast, the diffusive PINNs exhibit significantly larger full-field errors even at $\mathrm{Ra}=10^3$, which increase further at $\mathrm{Ra}=10^4$, reaching 88.7\%, 90.7\%, and 95.8\% for $u$, $v$, and $\theta$, respectively. These results demonstrate that the diffusive scaling fails to reproduce the DNS fields accurately within the prescribed training budget, whereas the free-fall scaling provides substantially more accurate predictions.

\begin{table}[t!]
\centering
\caption{Full-field relative $L_2$ errors in the velocity components and temperature against DNS, evaluated over the entire $64\times64$ computational grid using Eq.~\ref{eq:fullfield_relativeerror}.}
\label{tab:fielderror}
\begin{tabular}{@{}llrrr@{}}
\toprule
Scaling & $\mathrm{Ra}$ & $u$ (\%) & $v$ (\%) & $\theta$ (\%) \\
\midrule
Diffusive & $10^3$ & 22.3 & 41.5 & 54.1 \\
Free-fall & $10^3$ & 1.0 & 1.3 & 0.1 \\
\addlinespace
Diffusive & $10^4$ & 88.7 & 90.7 & 95.8 \\
Free-fall & $10^4$ & 1.0 & 1.2 & 0.9 \\
\addlinespace
Diffusive & $10^5$ & \texttt{NaN} & \texttt{NaN} & \texttt{NaN} \\
Free-fall & $10^5$ & 20.9 & 14.2 & 9.7 \\
\bottomrule
\end{tabular}
\end{table}

\begin{figure}[t!]
\centering
\includegraphics[width=\textwidth]{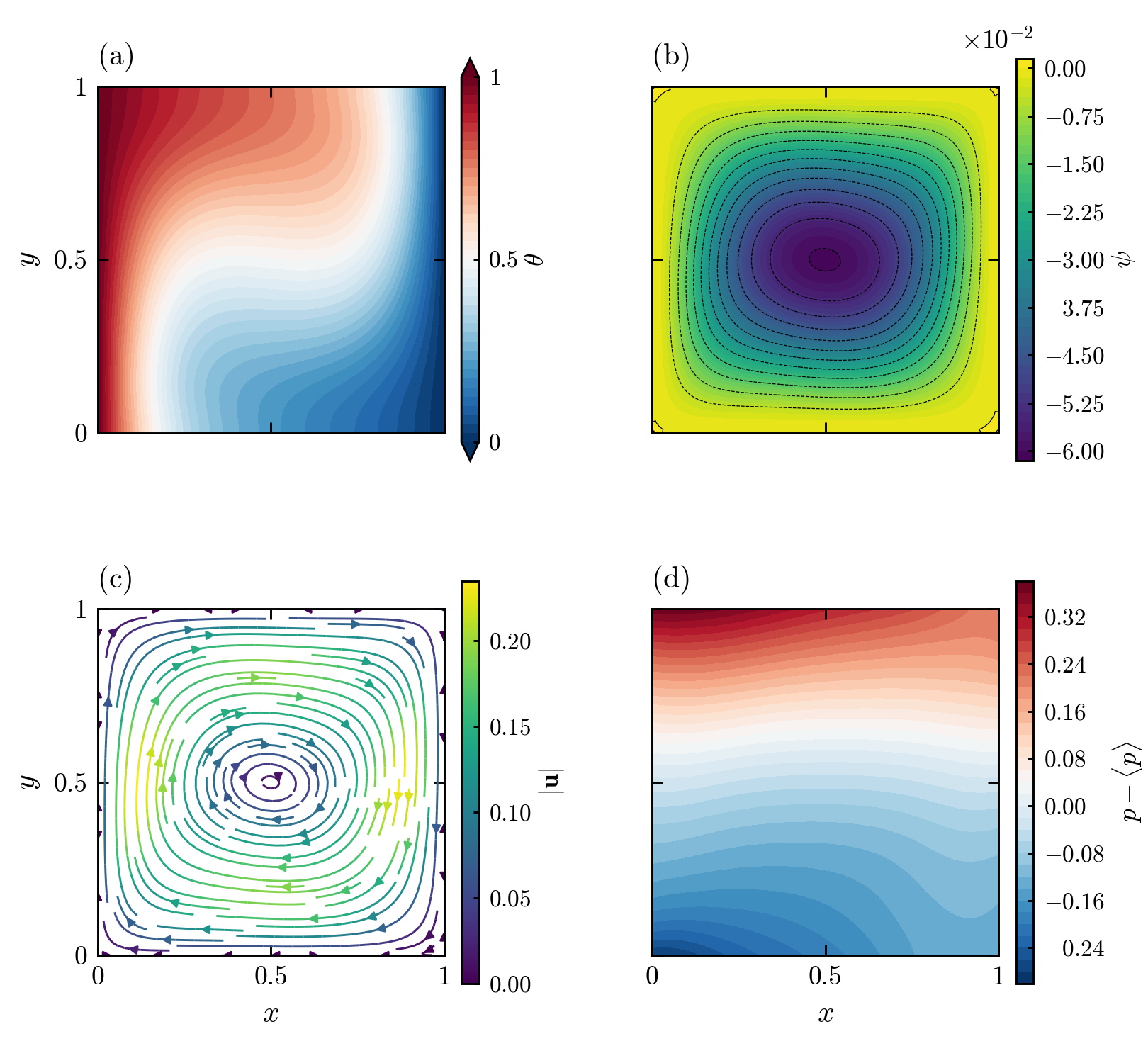}
\caption{Flow fields predicted by the free-fall PINN at $\mathrm{Ra}=10^4$: (a) temperature, (b) streamfunction, (c) velocity streamlines colored by speed, and (d) pressure relative to its spatial mean. Velocity, streamfunction, and pressure are expressed using the free-fall scaling.}
\label{fig:fields-freefall}
\end{figure}

\begin{figure}[t!]
\centering
\includegraphics[width=\textwidth]{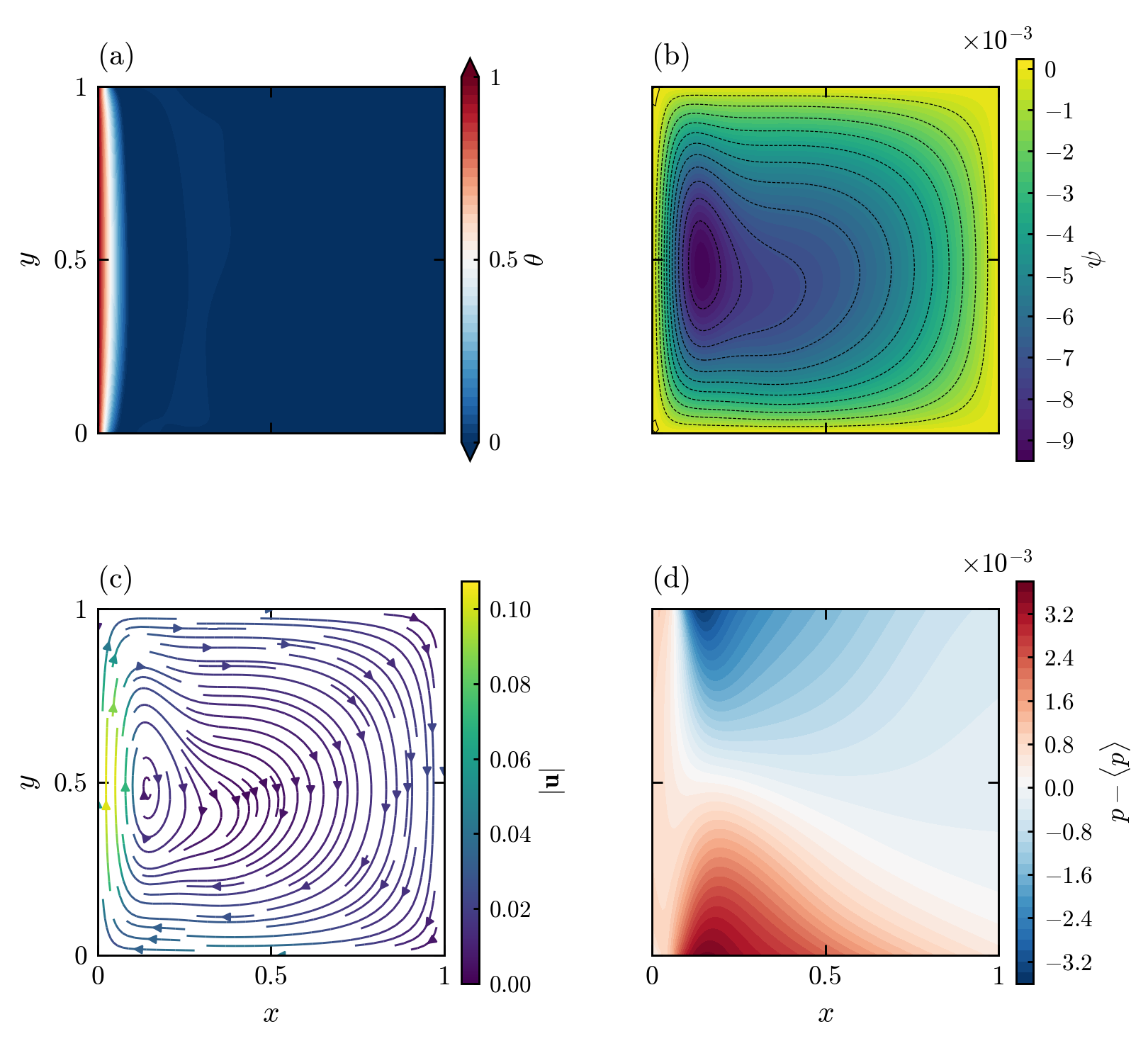}
\caption{Flow fields predicted by the diffusive PINN at $\mathrm{Ra}=10^4$, using the same network architecture and training budget as in Fig.~\ref{fig:fields-freefall}: (a) temperature, (b) streamfunction, (c) velocity streamlines colored by speed, and (d) pressure relative to its spatial mean. The velocity, streamfunction, and pressure fields are converted to the free-fall scale for comparison. The predicted temperature variations are largely confined to a thin near-wall region, in contrast to the smooth distribution obtained with the free-fall PINN.}
\label{fig:fields-diffusive}
\end{figure}

To further examine the differences in the predicted flow structures, Figs.~\ref{fig:fields-freefall} and \ref{fig:fields-diffusive} present the temperature, streamfunction, velocity streamlines, and pressure fields obtained using the free-fall and diffusive scalings, respectively, at $\mathrm{Ra}=10^4$. The free-fall PINN produces a smooth temperature distribution and a coherent circulation pattern, whereas the diffusive PINN exhibits a severely distorted temperature field, with variations largely confined to a thin near-wall region. These qualitative differences are consistent with the substantially larger errors observed for the diffusive scaling.

\begin{figure}[t!]
\centering
\includegraphics[width=\textwidth]{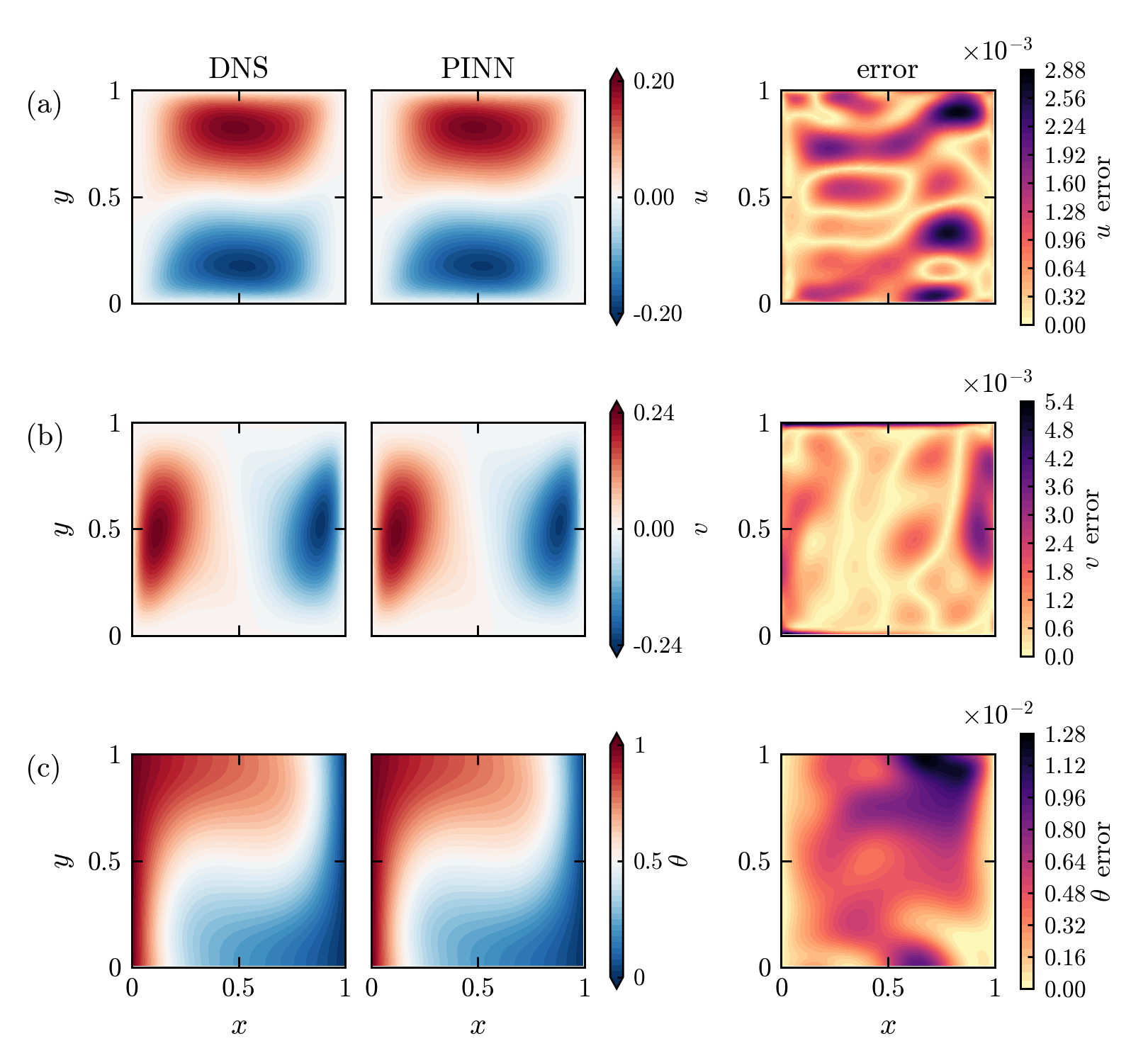}
\caption{Comparison of the free-fall PINN predictions with DNS at $\mathrm{Ra}=10^4$ for (a) horizontal velocity $u$, (b) vertical velocity $v$, and (c) temperature $\theta$. The columns show the DNS reference fields, PINN predictions, and absolute pointwise errors $|\phi_{\mathrm{PINN}}-\phi_{\mathrm{DNS}}|$, respectively. The DNS and PINN fields share the same color scale for each variable.}
\label{fig:dns-freefall}
\end{figure}

\begin{figure}[htbp]
\centering
\includegraphics[width=\textwidth]{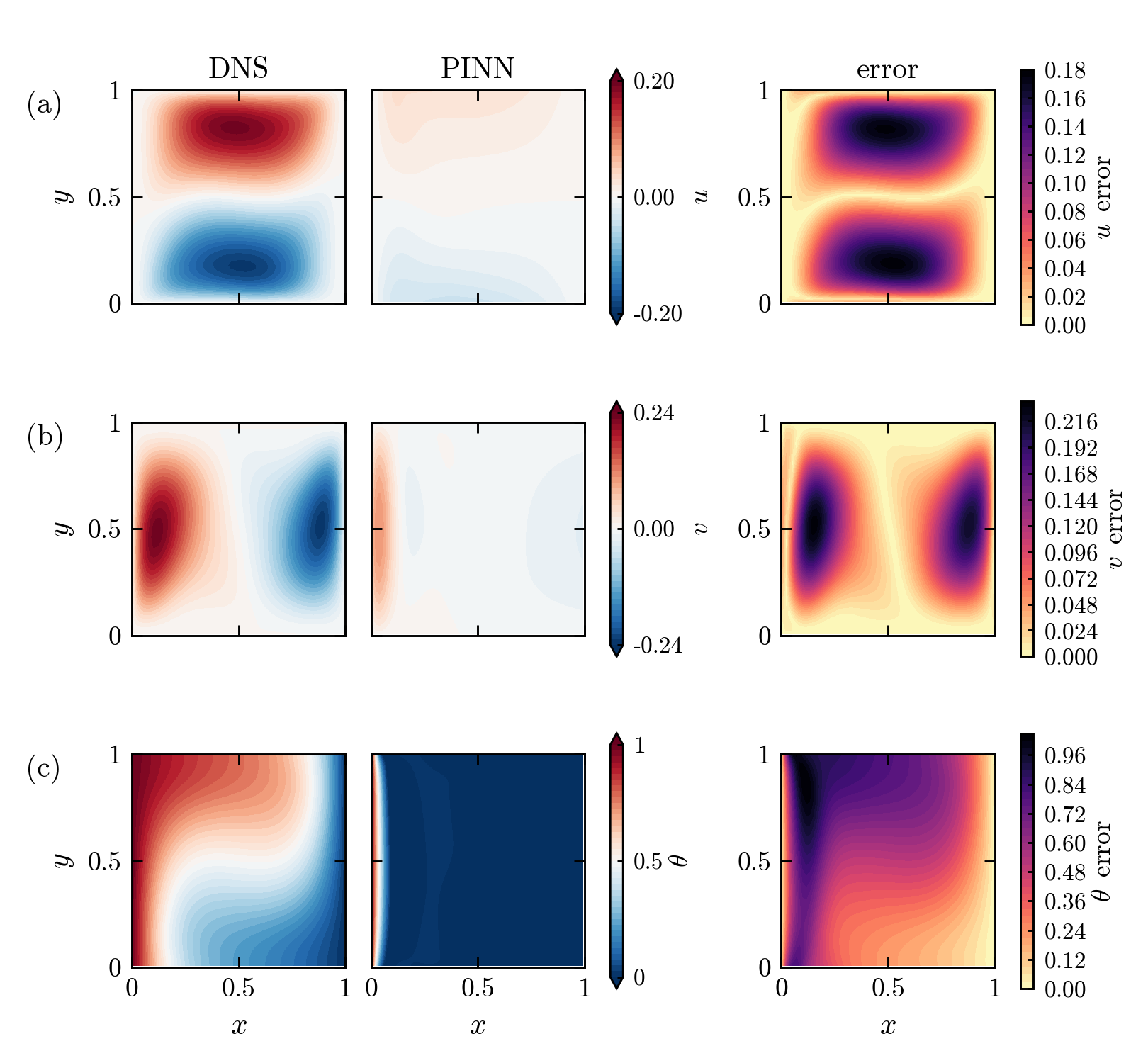}
\caption{Comparison of the diffusive PINN predictions with DNS at $\mathrm{Ra}=10^4$, following the same arrangement as Fig.~\ref{fig:dns-freefall}. The DNS and PINN fields share the same color scale for each variable, highlighting the substantial differences in field magnitudes and spatial distributions.}
\label{fig:dns-diffusive}
\end{figure}

A direct comparison of the predicted velocity and temperature fields with DNS is shown in Figs.~\ref{fig:dns-freefall} and \ref{fig:dns-diffusive}. The free-fall PINN closely reproduces the DNS fields, with small pointwise differences across the computational domain. In contrast, the diffusive PINN fails to capture the spatial distributions and magnitudes of the reference fields. These comparisons further confirm that the free-fall scaling provides substantially better accuracy and training convergence than the diffusive scaling under identical network architectures and training budgets.

\subsection{Flow structure comparison}

The differentially heated cavity (DHC) is characterized by a well-known recirculating flow pattern and a gently S-shaped temperature distribution at low Rayleigh numbers. Here, we compare the flow structures predicted by the diffusive and free-fall PINNs at $\mathrm{Ra}=10^4$.

Figure~\ref{fig:fields-freefall} shows the flow fields predicted by the free-fall PINN at $\mathrm{Ra}=10^4$. The temperature field exhibits the characteristic S-shaped stratification, while the streamfunction and velocity fields reveal a smooth, coherent recirculating cell. In contrast, the diffusive PINN fails to reproduce these characteristic flow features, as shown in Fig.~\ref{fig:fields-diffusive}. In particular, the temperature variations are confined to a thin near-wall region, with the bulk temperature remaining nearly uniform at $\theta\approx0$. Furthermore, the recirculating core is displaced from the cavity center, as evident from the streamfunction and velocity fields. The pressure field also differs considerably between the two scalings. While the free-fall PINN produces a smooth pressure distribution, the diffusive PINN exhibits two distinct pressure peaks near the left wall.

A direct comparison of the PINN predictions with DNS at $\mathrm{Ra}=10^4$ is presented in Figs.~\ref{fig:dns-freefall} and \ref{fig:dns-diffusive}. The free-fall PINN closely reproduces the DNS results for all three fields, with nearly indistinguishable spatial distributions. The pointwise errors are generally more than two orders of magnitude smaller than the characteristic field magnitudes, indicating excellent agreement with DNS. In contrast, the diffusive PINN fails to capture the characteristic flow structures. Although weak velocity features are observed near the cavity walls, the predicted velocity magnitudes are approximately six times smaller than those obtained from DNS. The temperature field exhibits the largest discrepancy, with no recognizable S-shaped stratification and pointwise errors exceeding 0.5 over much of the domain.

These comparisons demonstrate that the differences between the two scalings extend beyond training convergence and global accuracy to the prediction of the underlying flow structures. Under identical network architectures and training budgets, the free-fall PINN successfully captures the characteristic flow features of the DHC, whereas the diffusive PINN produces substantially distorted velocity, temperature, and pressure fields.

\subsection{Mechanism: amplification of gradient imbalance}
\label{sec:discussion}

Previous studies have employed non-dimensionalizations in which the buoyancy coefficient satisfies $B\approx O(1)$ \citep{cai2021heat,volk_pinn_2025}. However, the rationale for this choice requires further examination, particularly because the coefficients associated with buoyancy and diffusion directly influence the PDE residuals and their contributions to the loss function. To investigate the mechanism underlying the training difficulties observed with diffusive scaling, we analyze the gradient contributions arising from the vertical momentum equation.

The $y$-momentum residual can be expressed as $\mathrm{mom}_v=R-B\theta$, where $R$ contains the advection, pressure-gradient, and viscous terms, while $B=\mathrm{Ra}\,\mathrm{Pr}$ for diffusive scaling and $B=1$ for free-fall scaling (Eqs.~\ref{eq:diff-momy} and \ref{eq:ff-momy}). Near initialization, the velocity components are small ($u,v\approx0$), while the temperature remains $O(1)$. Assuming that $R$ is small compared with the buoyancy term, the residual can be approximated as $\mathrm{mom}_v\approx-B\theta$. For the corresponding squared residual loss, $\mathcal{L}_v=\mathrm{mom}_v^2$, the gradient with respect to a network parameter $w$ becomes

\begin{align}
\frac{\partial\mathcal{L}_v}{\partial w}
&=2\,\mathrm{mom}_v\,\frac{\partial\mathrm{mom}_v}{\partial w}\nonumber\\
&\approx 2(-B\theta)(-B)\frac{\partial\theta}{\partial w}
=2B^2\theta\frac{\partial\theta}{\partial w}.
\label{eq:gradscaling}
\end{align}

Thus, the buoyancy coefficient enters the gradient twice: once through the residual magnitude and again through its derivative with respect to the network parameters. Consequently, the temperature-dependent contribution to the gradient scales as $B^2$. For diffusive scaling, where $B=\mathrm{Ra}\,\mathrm{Pr}$, this amplification factor increases by two orders of magnitude for every decade increase in $Ra$, as shown in Table~\ref{tab:scalinglaw}. This trend is consistent with the deterioration in training accuracy reported in Table~\ref{tab:dns}, where the diffusive PINNs exhibit large but finite errors at $\mathrm{Ra}=10^3$ and $10^4$, followed by divergence at $\mathrm{Ra}=10^5$. In contrast, the free-fall scaling maintains $B=1$ for all $Ra$, thereby avoiding this particular source of gradient amplification. Although the viscous and thermal diffusion coefficients decrease with increasing $Ra$ (Table~\ref{tab:coefficients}), leading to thinner boundary layers and potentially greater spatial resolution requirements, they do not introduce the same $B^2$ amplification mechanism.

\begin{table}[t!]
\centering
\caption{Gradient-amplification factor $B^2$ estimated from Eq.~\eqref{eq:gradscaling} for diffusive scaling ($B=\mathrm{Ra}\,\mathrm{Pr}$), together with the corresponding PINN errors in the mean Nusselt number. The amplification factor increases by two orders of magnitude for every decade increase in $Ra$, consistent with the observed deterioration in training performance.}
\label{tab:scalinglaw}
\begin{tabular}{@{}lrrl@{}}
\toprule
$\mathrm{Ra}$ & $B=\mathrm{Ra}\,\mathrm{Pr}$ & $B^2$ & Diffusive $\overline{\mathrm{Nu}}$ error \\
\midrule
$10^3$ & 710 & $5.0\times10^5$ & 157.7\% \\
$10^4$ & 7{,}100 & $5.0\times10^7$ & 648.5\% \\
$10^5$ & 71{,}000 & $5.0\times10^9$ & \texttt{NaN} (diverged) \\
\bottomrule
\end{tabular}
\end{table}

\begin{figure}[t!]
\centering
\includegraphics[width=\textwidth]{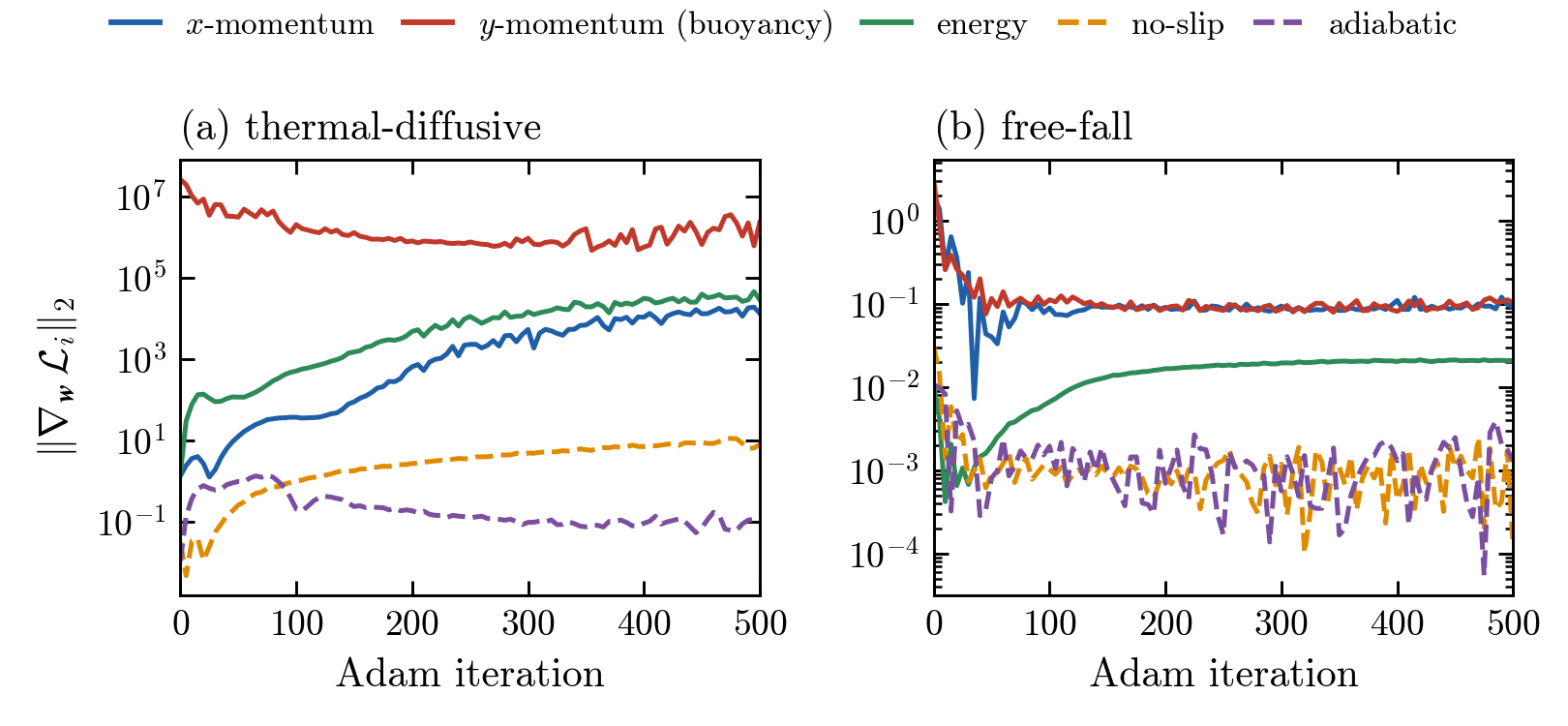}
\caption{$L_2$ norms of the gradients with respect to all network parameters for five raw, unweighted loss terms at $\mathrm{Ra}=10^4$, monitored over 500 Adam iterations using a common random seed: (a) diffusive scaling and (b) free-fall scaling. Note the different vertical scales. The diffusive $y$-momentum gradient remains above $10^5$, whereas the free-fall gradients lie approximately within $10^{-4}$--$10^0$.}
\label{fig:gradnorm}
\end{figure}

To examine this mechanism directly, we analyze the gradients of the individual loss terms at a fixed $\mathrm{Ra}=10^4$. Both PINNs are initialized using the same random seed and network architecture. Figure~\ref{fig:gradnorm} shows the $L_2$ norms of the gradients with respect to the network parameters for five raw, unweighted loss terms: the two momentum residuals, energy residual, no-slip boundary condition, and adiabatic boundary condition. These gradients are monitored over 500 Adam iterations. For the diffusive PINN, the gradient norm associated with the $y$-momentum residual is initially $8.35\times10^8$ times larger than that of the no-slip boundary condition and remains 5--7 orders of magnitude larger than the boundary-condition gradients throughout the training, as shown in Fig.~\ref{fig:gradnorm}(a). This indicates that the gradient imbalance persists rather than diminishing during the initial stages of optimization. 

In contrast, the free-fall PINN exhibits a much smaller initial gradient ratio, of order $10$, consistent with the absence of the large $B^2$ factor (Fig.~\ref{fig:gradnorm}(b)). During the first 100 iterations, the gradients associated with both momentum equations decrease substantially and approach comparable magnitudes, while those of the remaining loss terms also decrease. These results indicate that free-fall scaling produces a more balanced distribution of gradient magnitudes across the competing loss terms, facilitating optimization.

It is important to note that the gradients shown in Fig.~\ref{fig:gradnorm} are computed from the raw, unweighted loss terms. In the diffusive PINN, the boundary-condition losses are already assigned weights of 20, yet the initial gradient imbalance spans several orders of magnitude. Although further adjustment of the loss weights could alter the relative gradient contributions, it would not eliminate the amplification factor originating from the buoyancy coefficient within the momentum residual. This motivates the use of a non-dimensionalization that maintains $B=O(1)$, as achieved by the free-fall scaling and other formulations employed in previous studies \citep{cai2021heat,volk_pinn_2025}.

\section{Evaluation of existing strategies for mitigating gradient imbalance}

We note that the effect of non-dimensionalization on training in the previous section is closely related to the gradient imbalance between competing loss terms previously investigated by Wang et al.\ \citep{wang2021}. In the present case, however, the imbalance is strongly influenced by the choice of non-dimensionalization, which introduces a large coefficient directly into the PDE residual before any loss weighting is applied. To examine whether commonly used mitigation strategies can overcome this difficulty, we consider two approaches: sparse-data supervision and gradient-norm-based adaptive loss weighting, as discussed in the next section.

For sparse-data supervision, we introduce 20 DNS reference points with a data-loss weight of 20. The additional supervision improves the predicted fields by only a few percentage points, while the original full-field errors remain approximately 80--95\% (Table~\ref{tab:data}). Similarly, the gradient-norm-based adaptive weighting strategy either produces negligible improvements or excessively suppresses the contribution of the buoyancy-dominated momentum residual (Table~\ref{tab:adaptive}). These results suggest that neither approach, under the tested configurations, is sufficient to overcome the strong gradient imbalance associated with diffusive scaling. While adaptive weighting redistributes the contributions of individual loss terms during optimization, it does not directly remove the large buoyancy coefficient from the governing residual. Likewise, the hard boundary constraints introduced in Sec.~\ref{sec:methods} prevent convergence toward certain trivial solutions encountered with soft boundary constraints, but they do not eliminate the training difficulties arising from the large $\mathrm{Ra}\,\mathrm{Pr}$ coefficient, as demonstrated by the errors in Table~\ref{tab:dns}.

In what follows, we first discuss the effect of sparse data assimilation and then adaptive loss weighting. 

\subsection{Effect of sparse data assimilation}
\label{sec:data-results}

Table~\ref{tab:data} presents the full-field relative $L_2$ errors and the hot-wall mean Nusselt number ($\overline{\mathrm{Nu}}$) for the six configurations described in Sec.~\ref{sec:data-setup}. The mean Nusselt number is obtained by integrating the local Nusselt number, $\mathrm{Nu}_{\mathrm{local}}(y)=-\partial\theta/\partial x|_{x=0}$, along the hot wall, with the temperature gradient evaluated using automatic differentiation. 

For the diffusive scaling, sparse data assimilation provides only marginal improvements in accuracy. Velocity-only supervision reduces the errors in $u$ and $v$ from 88.7\% and 90.7\% to 82.0\% and 83.2\%, respectively, while the temperature error remains high at 95.0\%. The corresponding error in $\overline{\mathrm{Nu}}$ also remains above 600\%. Including temperature supervision ($\tau=1$) does not yield further improvement, with the temperature error increasing slightly to 95.3\%, despite direct supervision at 20 points. In contrast, the free-fall PINN benefits from sparse data assimilation. The temperature error decreases from 0.92\% without data to 0.41\% with velocity-only supervision ($\tau=0$) and further to 0.26\% when temperature data are included ($\tau=1$). Similar improvements are observed in the velocity components and $\overline{\mathrm{Nu}}$. These results indicate that sparse supervision can further improve an already accurate PINN solution but is insufficient to overcome the training difficulties encountered with diffusive scaling under the present conditions.

\begin{table}[t!]
\centering
\caption{Full-field relative $L_2$ errors against DNS and hot-wall mean Nusselt number at $\mathrm{Ra}=10^4$ for three data conditions: no data, sparse velocity supervision ($\tau=0$), and sparse velocity and temperature supervision ($\tau=1$). The supervised cases use 20 data points located along $x\approx0.5$.}
\label{tab:data}
\begin{tabular}{@{}llrrrrr@{}}
\toprule
Scaling & Data condition & $u$ (\%) & $v$ (\%) & $\theta$ (\%) & $\overline{\mathrm{Nu}}$ & $\overline{\mathrm{Nu}}$ err.\ (\%) \\
\midrule
Diffusive & No data & 88.7 & 90.7 & 95.8 & 17.13 & 672 \\
Diffusive & $\tau=0$ ($u,v$) & 82.0 & 83.2 & 95.0 & 15.95 & 619 \\
Diffusive & $\tau=1$ ($u,v,\theta$) & 82.5 & 84.3 & 95.3 & 15.85 & 614 \\
\addlinespace
Free-fall & No data & 0.98 & 1.24 & 0.92 & 2.272 & 2.40 \\
Free-fall & $\tau=0$ ($u,v$) & 0.61 & 1.16 & 0.41 & 2.252 & 1.53 \\
Free-fall & $\tau=1$ ($u,v,\theta$) & 0.57 & 1.06 & 0.26 & 2.250 & 1.41 \\
\bottomrule
\end{tabular}
\end{table}
\begin{figure}[t!]
\centering
\includegraphics[width=\textwidth]{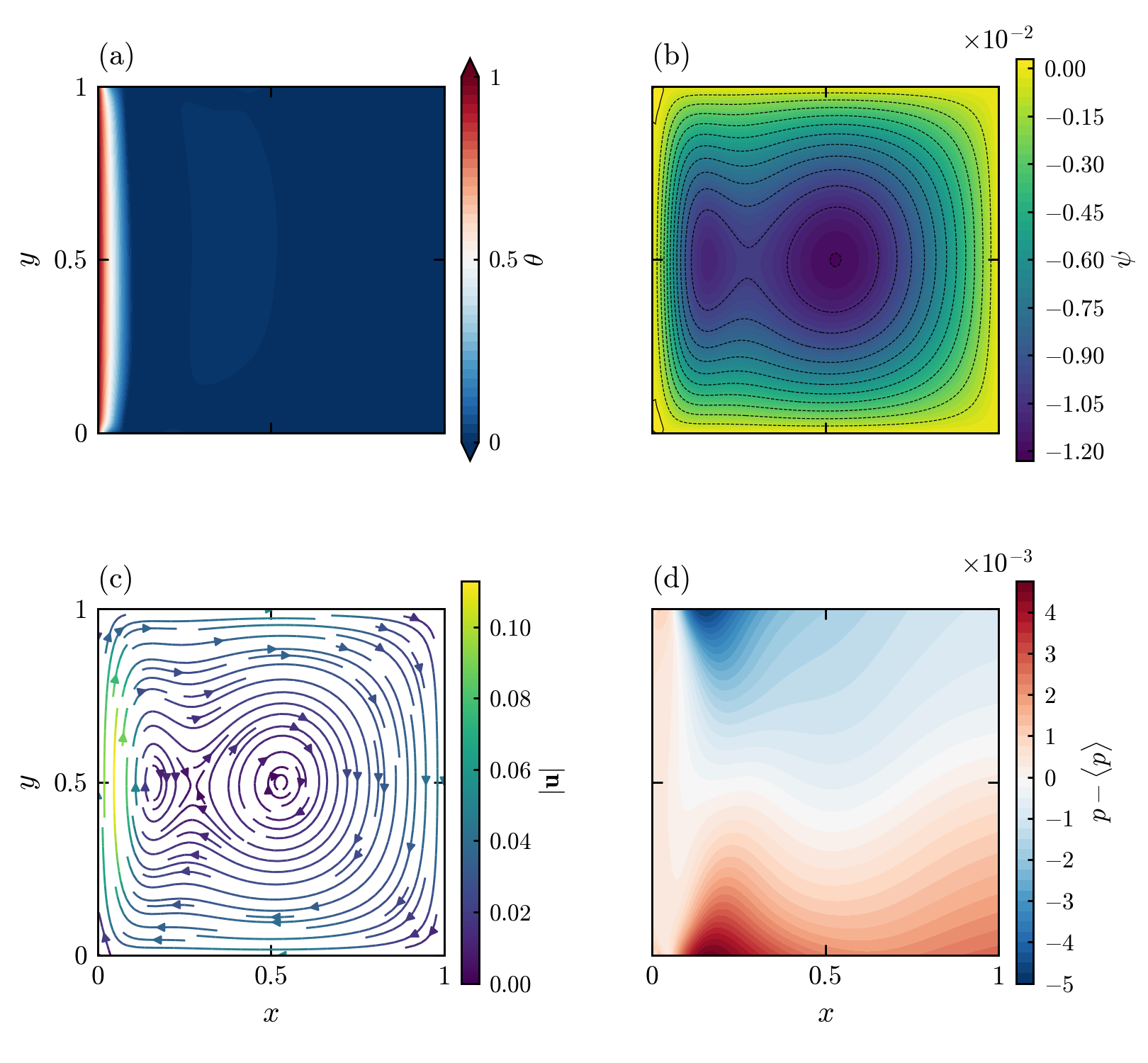}
\caption{Flow fields predicted by the diffusive PINN with sparse velocity supervision ($\tau=0$) at $\mathrm{Ra}=10^4$: (a) temperature, (b) streamfunction, (c) velocity streamlines colored by speed, and (d) pressure relative to its spatial mean. The velocity, streamfunction, and pressure fields are expressed using the free-fall scaling. Compared with the no-data case (Fig.~\ref{fig:fields-diffusive}), sparse supervision introduces an additional, spurious recirculating region.}
\label{fig:data-fields}
\end{figure}

\begin{figure}[t!]
\centering
\includegraphics[width=\textwidth]{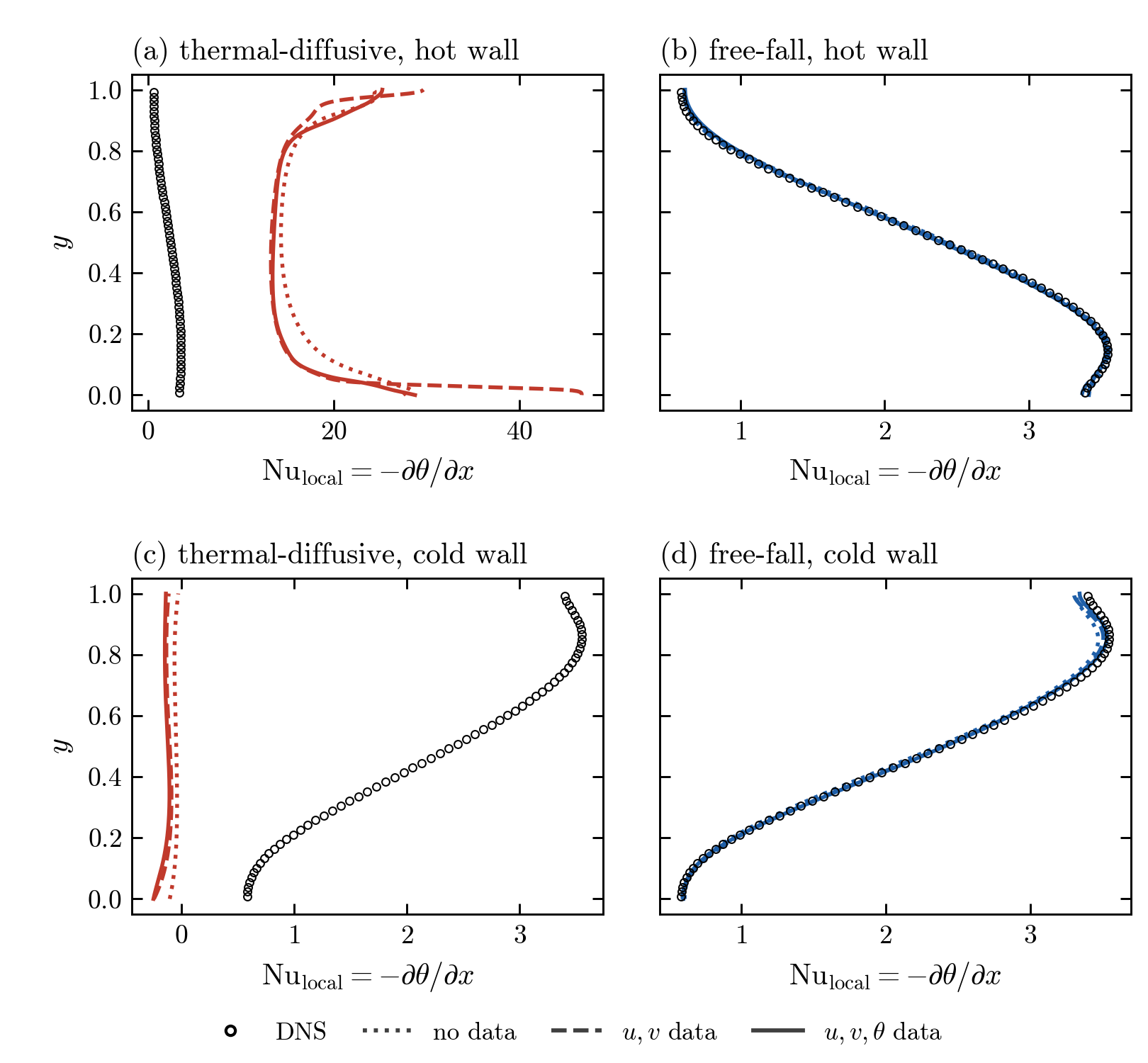}
\caption{Comparison of local Nusselt number distributions, $\mathrm{Nu}_{\mathrm{local}}(y)=-\partial\theta/\partial x$, at the hot wall ($x=0$, top row) and cold wall ($x=1$, bottom row) for $\mathrm{Ra}=10^4$. DNS results (symbols) are compared with PINN predictions under three data conditions: no data (dotted lines), velocity-only supervision ($u,v$, dashed lines), and velocity and temperature supervision ($u,v,\theta$, solid lines), for both diffusive and free-fall scalings.}
\label{fig:nusselt-data}
\end{figure}
Although sparse data assimilation slightly reduces the velocity errors in the diffusive PINN, it also alters the predicted flow topology. Figure~\ref{fig:data-fields} shows the flow fields obtained with velocity-only supervision ($\tau=0$). The temperature distribution remains largely unchanged from the no-data case, with variations confined to a thin near-wall region (Fig.~\ref{fig:fields-diffusive}(a)). However, the streamfunction now exhibits two distinct recirculating regions, with a smaller cell near $x\approx0.2$ and a larger cell near $x\approx0.55$, instead of the single off-center recirculating cell observed without supervision. This additional flow structure is inconsistent with the expected flow topology at $\mathrm{Ra}=10^4$. Wan et al.\ \citep{wan2001} reported the emergence of two counter-rotating eddies at approximately $\mathrm{Ra}=10^5$. The appearance of a second recirculating region at the lower Rayleigh number therefore suggests that sparse supervision modifies the predicted velocity field without recovering the correct overall flow structure. Consequently, the modest reduction in velocity errors does not necessarily correspond to an improvement in the physical consistency of the solution.

The limited effectiveness of sparse data assimilation is further demonstrated by the local Nusselt number distributions at the hot ($x=0$) and cold ($x=1$) walls, as shown in Fig.~\ref{fig:nusselt-data}. At the hot wall, all three diffusive PINNs (without data, with $u,v$ only, and with $u,v,\theta$) substantially over-predict the local Nusselt number, with values approximately 5--15 times larger than DNS. Furthermore, the predicted distributions exhibit nearly flat profiles in the central region and sharp peaks near both ends of the wall, in contrast to the smooth, monotonically decreasing DNS profile. These distributions remain nearly unchanged with the addition of sparse data. At the cold wall, the diffusive PINNs predict local Nusselt numbers close to zero, whereas the DNS values decrease smoothly from approximately 3.5 to 0.6. In contrast, the free-fall PINNs closely reproduce the DNS distributions at both walls for all three data conditions.

The discrepancy between the hot- and cold-wall heat transfer rates also reveals a significant inconsistency in the diffusive PINN solutions. For DNS, the mean Nusselt numbers at the hot and cold walls are 2.2184 and 2.2183, respectively, satisfying the steady-state heat balance to within approximately 0.005\%. The free-fall PINNs also maintain reasonably consistent heat transfer rates at both walls, with mean Nusselt numbers ranging from approximately 2.21 to 2.27. In contrast, the diffusive PINNs predict a mean Nusselt number of approximately 17 at the hot wall and nearly zero at the cold wall. This large discrepancy indicates that the predicted temperature fields do not adequately satisfy the global steady-state heat balance, despite the enforcement of the governing equations through the PINN loss function. Overall, these results demonstrate that sparse data assimilation, under the tested configurations, does not resolve the substantial inaccuracies and physical inconsistencies associated with diffusive scaling.

\subsection{Effect of adaptive loss weighting}
\label{sec:adaptive-results}
Next, we evaluate how adaptive loss weighting strategy impacts the training and PINN results. Table~\ref{tab:adaptive} presents the results of the five Adam-only training configurations described in Sec.~\ref{sec:adaptive-setup}. We find that neither adaptive weighting strategy improves the accuracy of the diffusive PINNs. The coarse-grained scheme produces velocity errors comparable to those of the fixed-weight baseline, with both $u$ and $v$ errors remaining approximately 100\%. Moreover, the mean Nusselt number ($\overline{\mathrm{Nu}}$) increases from 29.26 to 40.69, deviating further from the DNS reference value of 2.218. The final adapted weights provide insight into this behavior. The weights associated with the velocity and temperature boundary conditions, $\lambda_{\mathrm{vel}}$ and $\lambda_\theta$, increase to $1.1\times10^{10}$ and $7.4\times10^{11}$, respectively, while the $y$-momentum residual remains unweighted. Consequently, the scheme strongly emphasizes boundary-condition enforcement without directly addressing the imbalance among the individual PDE residuals. These results indicate that treating the PDE residuals as a single group is insufficient to mitigate the gradient imbalance observed in the diffusive PINN.

\begin{table}[t!]
\centering
\caption{Comparison of fixed and adaptive loss weighting strategies at $\mathrm{Ra}=10^4$, showing the full-field relative $L_2$ errors against DNS, hot-wall mean Nusselt number ($\overline{\mathrm{Nu}}$), and final adapted weight of the $y$-momentum residual. All configurations are trained using Adam for 5,000 iterations. The fixed-weight cases serve as matched-budget controls; therefore, the results are not directly comparable with Tables~\ref{tab:dns}--\ref{tab:data}. The DNS reference value is $\overline{\mathrm{Nu}}=2.218$, while $\overline{\mathrm{Nu}}=1$ corresponds to the pure-conduction limit.}
\label{tab:adaptive}
\begin{tabular}{@{}llrrrrr@{}}
\toprule
Scaling & Weighting & $u$ (\%) & $v$ (\%) & $\theta$ (\%) & $\overline{\mathrm{Nu}}$ & final $\lambda_{\mathrm{mom}_v}$ \\
\midrule
Diffusive & Fixed (control) & 99.7 & 104.5 & 100.1 & 29.26 & 1 \\
Diffusive & Adaptive, coarse & 100.0 & 100.0 & 97.5 & 40.69 & 1 \\
Diffusive & Adaptive, fine & 100.0 & 100.0 & 89.9 & 9.97 & $7.8\times10^{-6}$ \\
\addlinespace
Free-fall & Fixed (control) & 11.0 & 15.1 & 16.7 & 2.807 & 1 \\
Free-fall & Adaptive, fine & 99.5 & 99.6 & 36.9 & 1.016 & $2.2\times10^{-3}$ \\
\bottomrule
\end{tabular}
\end{table}

The fine-grained adaptive weighting scheme allows the individual PDE residuals to be weighted independently, but it also fails to recover an accurate diffusive PINN solution. In this case, the weight associated with the $y$-momentum residual decreases from 1 to $7.8\times10^{-6}$, while the corresponding raw residual remains $O(10^6)$. Thus, the apparent reduction in the weighted contribution of this term is primarily achieved by suppressing its weight rather than reducing the underlying residual. Although the temperature error decreases from 100.1\% to 89.9\% and $\overline{\mathrm{Nu}}$ improves from 29.26 to 9.97, the velocity errors remain approximately 100\%, indicating that the predicted solution is still far from the DNS reference.

More importantly, the fine-grained scheme also substantially deteriorates the accuracy of the free-fall PINN, which otherwise exhibits better convergence under fixed weighting. The velocity errors increase from 11.0\% and 15.1\% in the fixed-weight case to 99.5\% and 99.6\%, respectively, while $\overline{\mathrm{Nu}}$ decreases from 2.807 to 1.016. The latter approaches the pure-conduction limit, $\overline{\mathrm{Nu}}=1$, corresponding to a motionless fluid with a linear temperature distribution, $\theta=1-x$. Together with the nearly 100\% velocity errors, this suggests that the adaptive weighting scheme drives the solution toward a conduction-dominated state rather than the expected convective flow. The final weight of the $y$-momentum residual, $\lambda_{\mathrm{mom}_v}=2.2\times10^{-3}$, provides a possible explanation for this behavior. Since buoyancy enters through the vertical momentum equation, substantially reducing its contribution to the loss function weakens the coupling between the temperature and velocity fields. Consequently, the optimizer can favor solutions with weak convection while placing less emphasis on satisfying the buoyancy-driven momentum balance.

These results highlight two important limitations of the tested adaptive weighting strategies. First, neither coarse- nor fine-grained weighting provides an effective alternative to a suitable non-dimensionalization under the present training conditions. Second, balancing loss terms solely according to gradient magnitudes can be counterproductive when a particular residual represents an essential physical coupling. In the present case, excessive down-weighting of the $y$-momentum residual weakens the enforcement of buoyancy-driven convection and can lead to physically inaccurate solutions. These findings are specific to the adaptive weighting formulations and training settings considered here, as discussed in Sec.~\ref{sec:adaptive-setup}. In particular, the suppression of the $y$-momentum residual arises in our fine-grained generalization, whereas the original published formulation keeps the PDE-residual weight fixed.

For buoyancy-driven flows and other multiphysics problems, these findings highlight the importance of non-dimensionalization in PINN training. The choice of characteristic scales should therefore be considered alongside the network architecture, optimizer, and loss weights. In particular, a free-fall-type scaling that keeps the governing coefficients of order unity or smaller can mitigate coefficient-induced gradient amplification without requiring modifications to the network architecture or additional training procedures.

\section{Time evolution of the flow field}
\label{sec:unsteady-results}

\begin{figure}[t!]
\centering
\includegraphics[width=\textwidth]{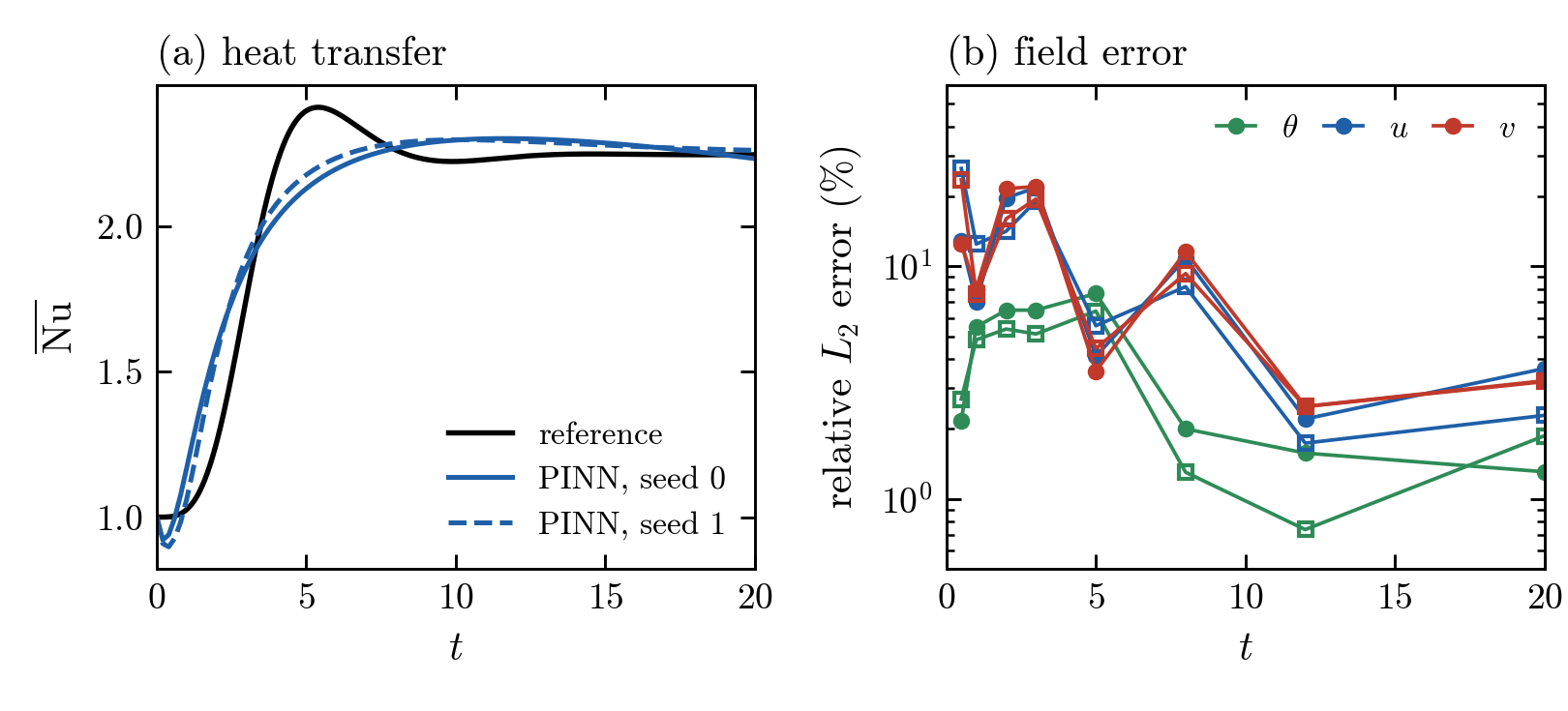}
\caption{Temporal evolution of buoyancy-driven flow from rest at $\mathrm{Ra}=10^4$ using free-fall scaling: (a) hot-wall mean Nusselt number, $\overline{\mathrm{Nu}}(t)$, comparing the reference solution with PINN predictions from two random seeds; (b) full-field relative $L_2$ errors in temperature ($\theta$) and velocity components ($u,v$) at the available reference snapshot times. Filled and open symbols represent seeds 0 and 1, respectively.}
\label{fig:unsteady-nu}
\end{figure}

In this section, we compare the time-evolution of Figure~\ref{fig:unsteady-nu}(a) compares the temporal evolution of the hot-wall mean Nusselt number, $\overline{\mathrm{Nu}}(t)$, predicted by the free-fall PINN with the reference solution. In the reference solution, $\overline{\mathrm{Nu}}$ initially equals 1, corresponding to the pure-conduction state, increases to a maximum of 2.408 at $t=5.4$, subsequently decreases to a minimum of 2.222 at $t=9.9$, and eventually approaches a steady value of 2.245. The PINN captures the initial and long-time behavior reasonably well, with $\overline{\mathrm{Nu}}(20)=2.232$ and 2.261 for the two random seeds, corresponding to deviations of $-0.6\%$ and $+0.7\%$ from the steady reference value, respectively. However, neither seed reproduces the transient overshoot observed in the reference solution. Instead, $\overline{\mathrm{Nu}}$ increases smoothly to approximately 2.30 at $t\approx10$--12 before gradually approaching its long-time value, thereby missing the transient oscillatory behavior. Furthermore, the PINN predicts a faster initial increase in $\overline{\mathrm{Nu}}$ than the reference, with maximum absolute errors of 0.334 and 0.302 for seeds 0 and 1, respectively, near $t\approx2$. These correspond to relative errors of approximately 23--27\%, as the reference $\overline{\mathrm{Nu}}$ is only about 1.2--1.3 at this time. Both seeds also exhibit an initial undershoot below the pure-conduction value, reaching approximately 0.92 and 0.90 at $t\approx0.3$, which is absent from the reference solution. The suppression of the transient overshoot and the smoother temporal evolution suggest that the PINN has difficulty resolving the relatively rapid changes during the initial development of convection.

\begin{figure}[t!]
\centering
\includegraphics[width=\textwidth]{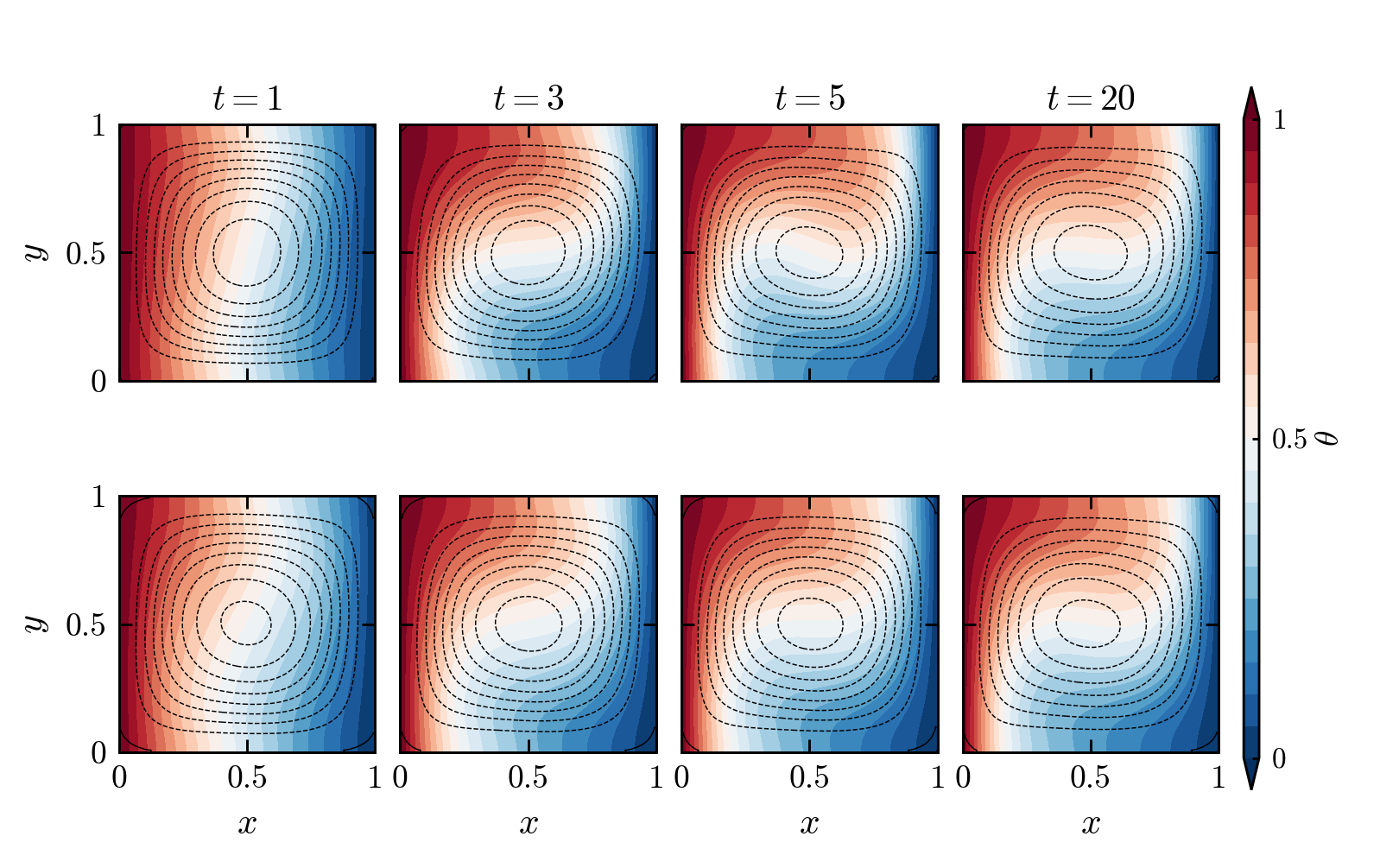}
\caption{Comparison of temperature fields (filled contours) and streamfunction contours (dashed lines) between the finite-difference reference (top row) and the free-fall PINN with seed 0 (bottom row) at $t=1$, 3, 5, and 20 for $\mathrm{Ra}=10^4$.}
\label{fig:unsteady-fields}
\end{figure}

The evolution of the full-field errors, shown in Fig.~\ref{fig:unsteady-nu}(b) and summarized in Table~\ref{tab:unsteady}, further highlights the differences between transient and long-time accuracy. During the initial transient ($t\le5$), the relative $L_2$ error in temperature remains approximately 2--8\%, decreasing to 1--2\% at $t=20$. The velocity errors are considerably larger, reaching 13--26\% at $t=0.5$ and remaining around 20\% at $t=2$--3, when the buoyancy-driven circulation develops and accelerates. However, these errors decrease substantially as the flow approaches its steady state, falling to approximately 2--4\% for $t\ge12$. Despite the relatively large transient errors, the PINN reproduces the overall spatial structures of the temperature and flow fields, as shown in Fig.~\ref{fig:unsteady-fields}. The predicted thermal fronts and recirculation patterns are qualitatively consistent with the reference at all selected times. The main discrepancies are associated with small displacements of the thermal boundary layers during the early transient and a slightly weaker circulation at $t=3$. By $t=20$, the predicted vortex core closely matches the reference in both location and shape.

\begin{table}[t!]
\centering
\caption{Comparison of free-fall PINN predictions from two random seeds with the finite-difference reference for unsteady flow from rest at $\mathrm{Ra}=10^4$. The final hot-wall mean Nusselt number, $\overline{\mathrm{Nu}}(20)$, is compared with the reference value of 2.245. The mean and maximum absolute Nusselt number errors, $|\Delta\overline{\mathrm{Nu}}|$, are evaluated over $0\le t\le20$. The remaining columns report full-field relative $L_2$ errors (\%) in the velocity components and temperature at $t=3$ and $t=20$.}
\label{tab:unsteady}
\small
\begin{tabular}{@{}lcccccccccc@{}}
\toprule
 & & \multicolumn{2}{c}{$|\Delta\overline{\mathrm{Nu}}|$} & \multicolumn{3}{c}{$t=3$} & \multicolumn{3}{c}{$t=20$} \\
\cmidrule(lr){3-4}\cmidrule(lr){5-7}\cmidrule(lr){8-10}
Seed & $\overline{\mathrm{Nu}}(20)$ & mean & max & $u$ & $v$ & $\theta$ & $u$ & $v$ & $\theta$ \\
\midrule
0 & 2.232 & 0.090 & 0.334 & 21.8 & 22.0 & 6.5 & 3.6 & 3.2 & 1.3 \\
1 & 2.261 & 0.079 & 0.302 & 19.1 & 19.5 & 5.1 & 2.3 & 3.2 & 1.9 \\
\bottomrule
\end{tabular}
\end{table}

Overall, these results demonstrate that the free-fall PINN can recover the long-time heat transfer characteristics and flow structures of buoyancy-driven convection from an initially stationary state without additional data supervision. However, its accuracy during the transient remains limited, particularly in capturing the overshoot in $\overline{\mathrm{Nu}}$ and the rapid development of the velocity field. The transient velocity errors are approximately an order of magnitude larger than those obtained near the steady state. Furthermore, compared with the steady PINN results at the same Rayleigh number (Table~\ref{tab:dns}), where the full-field errors remain around 1\%, the unsteady predictions exhibit substantially larger errors under the tested training configuration. This indicates that accurately resolving the temporal evolution of buoyancy-driven flow presents additional challenges, even when a suitable non-dimensionalization is employed. 

\section{Conclusions}
\label{sec:conclusion}
The present study investigated the influence of non-dimensionalization on the trainability and accuracy of physics-informed neural networks (PINNs) for buoyancy-driven flow in a differentially heated square cavity at $\mathrm{Pr}=0.71$ and $\mathrm{Ra}=10^3$--$10^5$. Two velocity scalings, thermal-diffusive and free-fall, were compared under identical network architectures, collocation-point budgets, optimizer settings, and training iterations to isolate the effect of non-dimensionalization.

In the thermal-diffusive scaling, commonly employed in classical CFD benchmarks, the buoyancy coefficient is $B=\mathrm{Ra}\,\mathrm{Pr}$, which increases by two orders of magnitude over the considered Rayleigh-number range. The corresponding PINNs exhibit errors of 157.7\% and 648.5\% in the mean Nusselt number ($\overline{\mathrm{Nu}}$) at $\mathrm{Ra}=10^3$ and $10^4$, respectively, and diverge at $\mathrm{Ra}=10^5$. In contrast, the free-fall scaling maintains a unit buoyancy coefficient ($B=1$) and exhibits substantially better training convergence and predictive accuracy. The free-fall PINNs agree with the reference DNS within 1.1\% in $\overline{\mathrm{Nu}}$ for $\mathrm{Ra}\le10^4$, while the error increases to 11.5\% at $\mathrm{Ra}=10^5$, with a maximum full-field relative $L_2$ error of 20.9\% in the horizontal velocity.

The observed training difficulties are associated with a strong coefficient-induced gradient imbalance. Near initialization, when the buoyancy contribution dominates the vertical momentum residual, the corresponding temperature-dependent loss gradient scales as $(\mathrm{Ra}\,\mathrm{Pr})^2$. Direct measurements of the individual loss gradients at $\mathrm{Ra}=10^4$ reveal that the gradient norm of the vertical momentum residual is initially $8.35\times10^8$ times larger than that of the no-slip boundary-condition loss in the diffusive formulation, compared with a much smaller ratio in the free-fall formulation. These results demonstrate how the coefficients introduced through non-dimensionalization can substantially influence the relative contributions of the governing residuals during optimization.

We further examined two commonly used strategies for mitigating gradient imbalance: sparse-data supervision and adaptive loss weighting. Neither strategy recovered accurate diffusive PINN solutions under the tested conditions. Supervision using 20 sparse DNS points, with velocity data alone or combined velocity and temperature data, left the full-field errors at approximately 82--96\% and introduced an additional, spurious recirculating region. Similarly, coarse-grained adaptive weighting failed to improve the solution, while fine-grained weighting excessively reduced the contribution of the vertical momentum residual. The latter also degraded the free-fall PINN predictions, driving the mean Nusselt number toward the pure-conduction limit ($\overline{\mathrm{Nu}}=1.016$). These findings indicate that adjusting the relative weights of loss terms does not necessarily overcome the gradient amplification originating from the governing PDE coefficients.

The free-fall formulation was also extended to unsteady convection developing from rest at $\mathrm{Ra}=10^4$, with initial conditions $u=v=0$ and $\theta=1-x$. For two random initializations, the predicted long-time mean Nusselt number agrees with the finite-difference reference within 0.7\%, and the overall flow structures are reproduced accurately. However, the PINNs fail to capture the transient overshoot in $\overline{\mathrm{Nu}}(t)$, predicting a peak of approximately 2.30 compared with the reference value of 2.41. The velocity errors remain around 20\% during the early transient before decreasing to approximately 2--4\% at $t=20$. These results suggest that, although free-fall scaling improves trainability and long-time accuracy, resolving transient flow dynamics remains challenging.

Overall, the findings establish non-dimensionalization as an important design consideration in PINN formulations for buoyancy-driven flows. Selecting characteristic scales that avoid excessively large PDE coefficients can mitigate gradient amplification at its source, rather than relying solely on auxiliary data or adaptive loss weighting. The present investigation is limited to $\mathrm{Ra}\le10^5$, a single network architecture, and one random initialization per steady configuration, while the unsteady analysis considers only $\mathrm{Ra}=10^4$ with free-fall scaling. Future work should examine higher Rayleigh numbers, different Prandtl numbers, additional network architectures and random initializations, direct comparisons of unsteady diffusive and free-fall formulations, improved resolution of transient dynamics, and extensions to three-dimensional buoyancy-driven flows.

\section*{Declaration of competing interest}
The authors declare that they have no known competing financial interests or personal relationships that could have appeared to influence the work reported in this paper.

\section*{Declaration of generative AI and AI-assisted technologies in the writing process}
During the preparation of this work the authors used Claude (Anthropic) for editing the manuscript. After using this tool, the authors reviewed and edited the content as needed and take full responsibility for the content of the publication.

\section*{Acknowledgements}

\section*{Data availability}
The training code, trained models, DNS reference data, and analysis scripts that support the findings of this study will be made available upon reasonable request.

\appendix

\section{Sparse-data assimilation setup}
\label{sec:data-setup}

In addition to the loss-balancing strategies discussed in the Introduction, sparse data assimilation provides an alternative approach to improving PINN training by incorporating direct supervision from reference data. Such data may be obtained from coarse numerical simulations or a limited number of experimental measurements. To evaluate the effectiveness of this approach, we introduce an additional data loss at $\mathrm{Ra}=10^4$, defined as

\begin{equation}
\mathcal{L}_{\mathrm{data}}
=
\mathrm{MSE}(u,u_{\mathrm{DNS}})
+
\mathrm{MSE}(v,v_{\mathrm{DNS}})
+
\tau\,\mathrm{MSE}(\theta,\theta_{\mathrm{DNS}}),
\end{equation}

where $\tau\in\{0,1\}$ controls the inclusion of temperature supervision. For $\tau=0$, only the velocity components are supervised, whereas $\tau=1$ includes both velocity and temperature data. This term is added to the total loss in Eq.~\eqref{eq:loss-total} with a weight of 20.

The reference data consist of 20 DNS points uniformly distributed in the vertical direction along the grid column nearest $x=0.5$. For the diffusive scaling, the DNS velocity components are converted from the free-fall scale by multiplying them by $\sqrt{\mathrm{Ra}\,\mathrm{Pr}}$. We verify that the trained diffusive PINN predicts velocities of $O(1$--$10)$ at the supervised locations, consistent with the scaled reference values. All other training settings are kept identical to those described in Sec.~\ref{sec:loss}.

A total of six configurations are considered, comprising two non-dimensionalizations (diffusive and free-fall) and three data conditions (no data, velocity-only supervision, and combined velocity and temperature supervision). This controlled comparison allows us to assess whether sparse data assimilation can improve the accuracy of PINNs trained under different non-dimensionalizations.

\section{Adaptive loss-weighting setup}
\label{sec:adaptive-setup}

The second strategy considered is adaptive loss weighting, motivated by the gradient-balancing approach of Wang et al.\ \citep{wang2021}. At $\mathrm{Ra}=10^4$, the fixed loss weights in Eq.~\eqref{eq:loss-total} are replaced by adaptive weights $\lambda_i$, which are updated every 10 training iterations based on the $L_2$ norms of the gradients of the individual, unweighted loss terms. The gradient norm associated with each loss component is defined as $g_i=\|\nabla_{\boldsymbol{w}}\mathcal{L}_i\|_2$, where $\boldsymbol{w}$ denotes the trainable network parameters. The weights are updated according to

\begin{equation}
\lambda_i
\leftarrow
(1-\alpha)\lambda_i
+
\alpha\frac{g_{\mathrm{ref}}}{g_i},
\qquad \alpha=0.1,
\label{eq:anneal}
\end{equation}

where $g_{\mathrm{ref}}$ is a reference gradient norm. This update aims to bring the weighted gradient magnitudes of the different loss components toward a common reference value.

We investigate two levels of loss decomposition. In the \emph{coarse-grained} scheme, the three PDE residuals are treated as a single group with a fixed weight of unity, and their combined gradient norm serves as the reference ($g_{\mathrm{ref}}=g_{\mathrm{pde}}$). Only the boundary-condition weights are adapted. This formulation follows the general structure of the method proposed by Wang et al.\ \citep{wang2021}. In the \emph{fine-grained} scheme, which is an extension introduced in the present study, the five loss components ($x$-momentum, $y$-momentum, energy, no-slip, and adiabatic boundary conditions) are assigned independent adaptive weights. The reference gradient norm is defined as the geometric mean of the five individual gradient norms.

This distinction is particularly relevant to the present problem because the gradient imbalance is not restricted to the PDE and boundary-condition groups. As demonstrated in Fig.~\ref{fig:gradnorm}, the vertical momentum residual produces substantially larger gradients than the other PDE residuals under diffusive scaling. Consequently, the coarse-grained scheme cannot directly address the imbalance among the individual governing equations, whereas the fine-grained scheme allows their relative contributions to be adjusted independently.

All adaptive weighting experiments are performed using 5,000 Adam iterations without subsequent L-BFGS optimization. This choice avoids the complications associated with changing loss weights during the line-search procedure of L-BFGS, which assumes a fixed objective function. To ensure a consistent comparison, fixed-weight baseline simulations are also performed using the same Adam-only training budget for both scalings. Therefore, the absolute errors obtained in these experiments should not be directly compared with those reported in Secs.~\ref{sec:results}--\ref{sec:data-results}, where L-BFGS optimization is also employed.

Three limitations of the adaptive weighting experiments should be noted. First, the present implementation uses global $L_2$ gradient norms rather than the layer-wise gradient statistics employed in the original formulation of Wang et al.\ \citep{wang2021}; therefore, it should not be regarded as an exact reproduction of that method. Second, the fine-grained scheme is a generalization introduced here, whereas the original approach retains a fixed weight for the PDE residual. Third, only one set of adaptation parameters ($\alpha=0.1$, with updates every 10 iterations) is considered. The conclusions regarding adaptive weighting are therefore restricted to the formulations and parameter settings investigated in this study.

\section{Unsteady formulation from rest}
\label{sec:unsteady-setup}

To investigate the applicability of the free-fall formulation to time-dependent convection, we consider the development of buoyancy-driven flow from an initially stationary state at $\mathrm{Ra}=10^4$. Only the free-fall scaling is examined, as the diffusive formulation exhibits substantial training difficulties even for the corresponding steady problem (Sec.~\ref{sec:results}). Time is non-dimensionalized using the free-fall time scale, $L/U$, where $U=\sqrt{g\beta\Delta\theta L}$. The unsteady governing equations are therefore

\begin{align}
u_t + u u_x + v u_y &= -p_x + \sqrt{\mathrm{Pr}/\mathrm{Ra}}\,(u_{xx}+u_{yy}), \label{eq:un-momx} \\
v_t + u v_x + v v_y &= -p_y + \sqrt{\mathrm{Pr}/\mathrm{Ra}}\,(v_{xx}+v_{yy}) + \theta, \label{eq:un-momy} \\
\theta_t + u \theta_x + v \theta_y &= \tfrac{1}{\sqrt{\mathrm{Ra}\,\mathrm{Pr}}}\,(\theta_{xx}+\theta_{yy}). \label{eq:un-energy}
\end{align}

The cavity is initially at rest with a linear conduction temperature profile, such that $u=v=0$ and $\theta=1-x$ at $t=0$. The network architecture is extended to include time as an additional input, resulting in three inputs $(x,y,t)$, four hidden layers with 64 neurons each, and three outputs $(\psi_{\mathrm{raw}},p_{\mathrm{raw}},\theta_{\mathrm{raw}})$. The time input is rescaled by the prescribed time horizon. The network employs $\tanh$ activation functions and contains 12,931 trainable parameters.

The initial conditions are enforced directly through the network outputs, together with the hard boundary constraints described in Sec.~\ref{sec:methods}, using the transformations

\begin{align}
\psi &= (1-e^{-t})\,x(1-x)\,y(1-y)\,\psi_{\mathrm{raw}}, \nonumber \\
\theta &= (1-x) + (1-e^{-t})\,x(1-x)\,\theta_{\mathrm{raw}}, \\
p &= (1-e^{-t})\,p_{\mathrm{raw}}. \nonumber
\end{align}

These transformations ensure that the initial velocity vanishes and the temperature satisfies $\theta=1-x$ at $t=0$, independently of the network parameters. They also enforce the no-penetration condition on all walls and the prescribed temperatures on the vertical walls throughout the simulation. The no-slip and adiabatic boundary conditions remain soft constraints, with the same loss weights as in Eq.~\eqref{eq:loss-total}, and are enforced using collocation points sampled over both space and time.

Training follows the causal weighting strategy proposed by Wang et al.\ \citep{wang2022causal}. The time interval $0\le t\le20$ is divided into $M=20$ consecutive bins of unit duration. During each training iteration, 300 interior collocation points and 15 points per wall are sampled within each time bin. Denoting the total loss associated with the $i$th bin by $\mathcal{L}_i$, the overall training loss is defined as

\begin{equation}
\mathcal{L}_{\mathrm{causal}}
=
\frac{1}{M}\sum_{i=1}^{M}w_i\mathcal{L}_i,
\end{equation}

where the causal weights are given by

\begin{equation}
w_1=1,\qquad
w_i=\exp\left(-\epsilon\sum_{k<i}\mathcal{L}_k\right),
\quad i>1.
\end{equation}

The cumulative loss appearing in the causal weights is detached from the computational graph during backpropagation. This formulation reduces the contribution of later time bins when the losses in preceding bins remain large, thereby encouraging the network to learn the temporal evolution progressively. The causal parameter is initialized at $\epsilon=1$, with a prescribed rule to double its value when the minimum causal weight exceeds 0.99. However, this condition is not reached during training, and $\epsilon$ remains unchanged.

The network is trained using 6,000 Adam iterations, with an initial learning rate of $10^{-3}$ reduced to zero through cosine annealing, followed by 1,000 L-BFGS iterations using a fixed set of collocation points. Two independent random initializations, corresponding to seeds 0 and 1, are considered to assess the sensitivity of the predictions to initialization.

The PINN predictions are compared against an independent finite-difference reference solution of the same governing equations in streamfunction--vorticity form. The reference solver employs second-order central differences on a uniform $129\times129$ grid, classical fourth-order Runge--Kutta time integration with $\Delta t=5\times10^{-4}$, a fast sine-transform Poisson solver for the streamfunction, and Thom's formula \citep{thom1933} for the wall vorticity. The simulation is initialized from the same stationary conduction state as the PINN.

The finite-difference solution approaches a steady mean Nusselt number of $\overline{\mathrm{Nu}}=2.2455$, consistent with the established benchmark value of approximately 2.24 \citep{devahldavis1983}. It also agrees closely with the steady DNS results of Sec.~\ref{sec:dns}, with full-field relative $L_2$ differences of 0.08\%, 0.8\%, and 1.1\% in $\theta$, $u$, and $v$, respectively. A grid-sensitivity comparison using a coarser $65\times65$ grid with $\Delta t=10^{-3}$ shows that the predicted $\overline{\mathrm{Nu}}(t)$ differs by at most 0.004 over $0\le t\le30$. These comparisons support the use of the finite-difference solution as a reference for evaluating the transient PINN predictions.

\bibliographystyle{elsarticle-num}
\bibliography{references_zotero,references}
\end{document}